\documentclass[twocolumn,article,
floatfix,superscriptaddress,longbibliography,nofootinbib]{revtex4-1}

\usepackage{amsmath}
\usepackage{dsfont}
\usepackage{verbatim}
\usepackage{braket}
\usepackage{titlesec}
\usepackage{graphicx}
\usepackage[usestackEOL]{stackengine}
\usepackage{array}
\usepackage{tikz}
\usepackage{xcolor}
\usepackage{epsfig} 
\usepackage{amssymb} 
\usepackage{longtable} 
\usepackage{float}
\usepackage{mathtools}
\usepackage{xparse}
\usepackage{times}
\usepackage[normalem]{ulem}

\renewcommand{\eqref}[1]{Eq. ({\ref{#1}})}

\newcommand{\OPc}[2]{\hat{#1}_{#2}^{\dag}}
\newcommand{\OP}[2]{\hat{#1}_{#2}^{\vphantom{\dag}}}
\newcommand{\CD}[1]{\OPc{c}{#1}}
\newcommand{\C}[1]{\OP{c}{#1}}

\begin{document}

\title{Light-induced switching between singlet and triplet superconducting states}

\author{Steven Gassner}
\email{sgassner@sas.upenn.edu}
\affiliation{Department of Physics and Astronomy, University of Pennsylvania, Philadelphia, PA 19104}
\author{Clara S. Weber}
\affiliation{Department of Physics and Astronomy, University of Pennsylvania, Philadelphia, PA 19104}
\affiliation{Institut f\"ur Theorie der Statistischen Physik, RWTH Aachen and JARA - Fundamentals of Future Information Technology, D-52056 Aachen,
Germany}
\author{Martin Claassen}
\email{claassen@sas.upenn.edu}
\affiliation{Department of Physics and Astronomy, University of Pennsylvania, Philadelphia, PA 19104}

\date{\today}

\begin{abstract}
While the search for topological triplet-pairing superconductivity has remained a challenge, recent developments in optically stabilizing metastable superconducting states suggest a new route to realizing this elusive phase. Here, we devise a testable theory of competing superconducting orders that permits ultrafast switching to an opposite-parity superconducting phase in centrosymmetric crystals with strong spin-orbit coupling. Using both microscopic and phenomenological models, we show that dynamical inversion symmetry breaking with a tailored light pulse can induce odd-parity (spin triplet) order parameter oscillations in a conventional even-parity (spin singlet) superconductor, which when driven strongly can send the system to a competing minimum in its free energy landscape. Our results provide new guiding principles for engineering unconventional electronic phases using light, suggesting a fundamentally non-equilibrium route toward realizing topological superconductivity.

\end{abstract}

\maketitle

\section*{Introduction}

Topological superconductors are elusive unconventional superconducting phases \cite{schnyder2008,qizhang2011,satoReview2017} that can host topologically-protected Majorana boundary modes and non-Abelian vortex excitations \cite{alicea2012,elliott2015}, which are of fundamental as well as tremendous practical interest as a route towards fault-tolerant quantum computing \cite{nayak2008}. Spin-triplet superconductors with finite angular momentum Cooper pairs \cite{Sigrist1991,Norman2011} have long been regarded as particularly promising candidates, with degeneracies between nodal order parameters expected to favor a chiral topological superconducting state \cite{kallinReview2016}. However, spin-triplet pairing remains rare in nature and signatures of chiral topological order remain inconclusive, despite several candidate compounds such as Sr$_2$RuO$_4$ \cite{Mackenzie2003, Maeno2012, Sharma2020} or UTe$_2$ \cite{Ran2019, Aoki2019} being placed under exceptional experimental scrutiny.

At the same time, a series of pioneering pump-probe experiments have established irradiation with light as an alternative and fundamentally non-equilibrium tool for interrogating and manipulating superconducting phases on ultrafast time scales, ranging from time-resolved probes of Higgs \cite{Anderson1958,Shimano2020,Matsunaga2013,Yuzbashyan2006} and Leggett \cite{Leggett1966,Huang2016,Zhao2020,Kamatani2022} mode oscillations in conventional and multi-gap superconductors to the light-induced enhancement or induction of long-lived superconducting signatures in the fullerides \cite{mitrano2016,Cavalleri2018,Budden2021}. With the underlying mechanisms still under substantial debate, these observations coincide with broader experimental \cite{Zhang2014, Beaud2014, Basov2017, Yoshikawa2021} and theoretical efforts \cite{claassen2019,Yu2021,sentef2017,Sun2020,Sun2021} in exploring thermal and non-thermal pathways to suppress or control competing ordered phases with light.

\begin{figure}[t]
    \centering
    \includegraphics[width=0.44\textwidth]{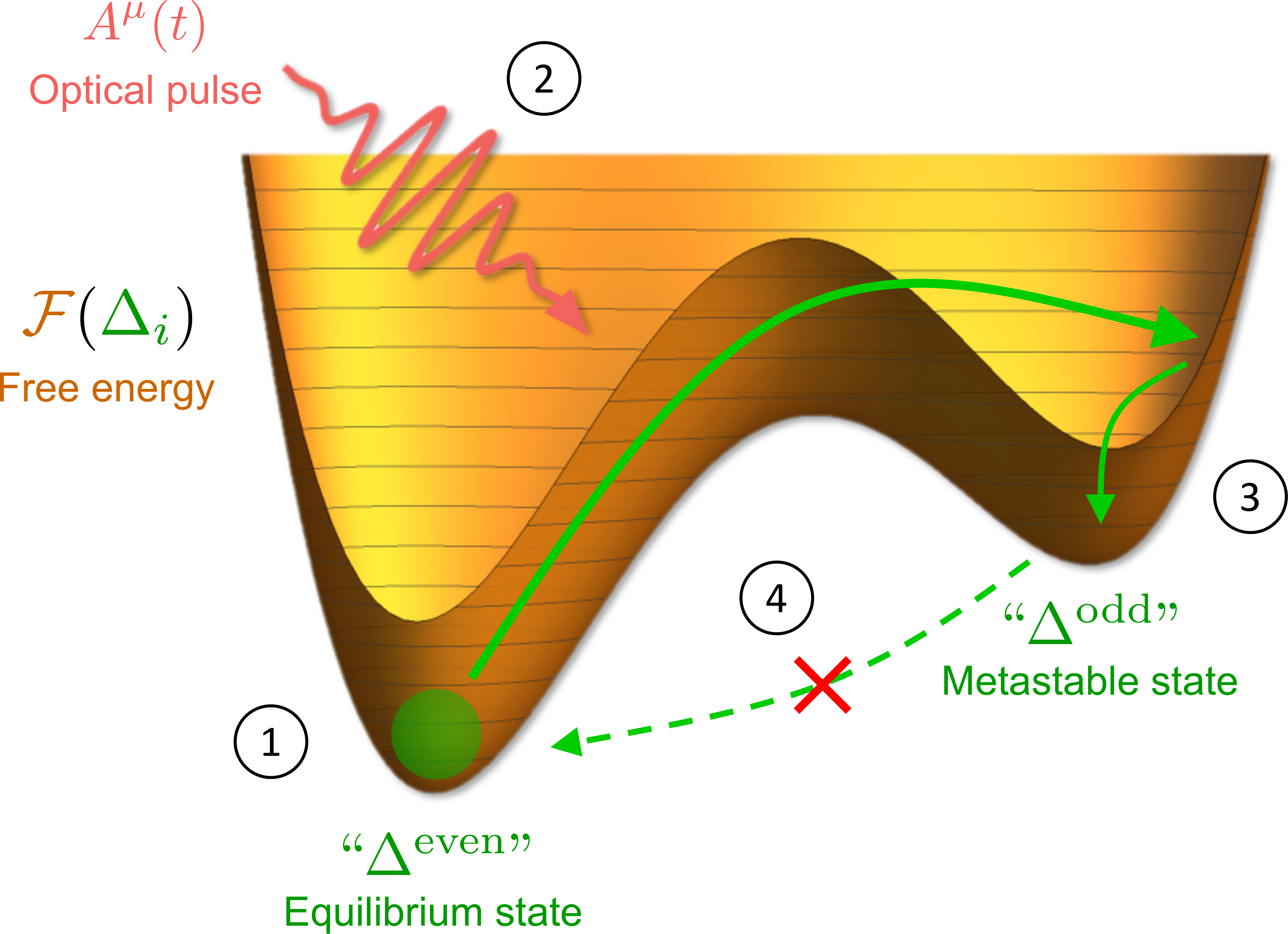}
    \caption{\textbf{Light-induced metastable triplet superconductors.} (1) The system starts in a superconducting phase that is even-parity (singlet-pairing) in equilibrium. (2) An optical pulse dynamically breaks inversion symmetry, driving the system toward a local minimum in its free energy landscape that comprises multiple competing order parameters. (3) The system relaxes into a metastable superconducting state that is odd-parity (triplet-pairing). (4) Since inversion symmetry is restored after the pulse, equilibration back to the opposite-parity state is suppressed.}
    \label{fig:Schematic}
\end{figure}

These results immediately raise the tantalizing question of whether elusive topological spin-triplet superconducting states can instead emerge as metastable phases upon irradiating a conventional superconductor with light. Consider an inversion-symmetric material with a conventional $s$-wave superconducting phase that preempts a closely competing topological spin-triplet pairing instability in equilibrium. In addition to the usual Higgs mode, such a system must necessarily retain additional amplitude modes in alternative pairing channels called Bardasis-Schrieffer (BS) modes \cite{bardasisschrieffer61,scalapino2009,Maiti2016}, which include odd-parity amplitude modes corresponding to spin-triplet pairing. These modes can lie below the gap in the case of closely-competing orders \cite{Lee2022} { provided that long-range Coulomb interactions do not push the mode into the pair-breaking continuum \cite{Hackner2023}}. While this mode remains decoupled in equilibrium from conventional Higgs oscillations in centrosymmetric materials, a tailored ultrafast light pulse can transiently break inversion symmetry and combine with strong spin-orbit coupling to induce odd-parity amplitude oscillations in a conventional superconductor. At a moderate fluence, the material can refuse to relax back to its equilibrium phase, instead becoming trapped in a metastable spin-triplet superconducting phase, with thermalization to the equilibrium spin-singlet phase suppressed due to restored inversion symmetry after the pulse.

In this work, we illustrate this mechanism as both a new route towards engineering spin-triplet and topological superconducting phases out of equilibrium as well as a generic probe of competing odd-parity instabilities in conventional superconductors. The protocol is summarized in Figure \ref{fig:Schematic}. We first identify ultrafast and two-color pulses as two complementary routes to dynamically break inversion symmetry, and demonstrate using a minimal model of quasiparticle dynamics that they can conspire with spin-orbit coupling to transiently induce odd-parity order parameter oscillations in a system with even-parity order. We then derive an effective time-dependent Ginzburg-Landau theory dictated via symmetry and present an explicit switching protocol for driving the system to settle into a metastable odd-parity superconducting state. Remarkably, we find that the coupling between equilibrium conventional $s$-wave order and a competing spin-triplet order parameter necessarily scales linearly with the field strength, in contrast to ordinary Higgs mode excitations which scale quadratically with the field. We illustrate the proposed mechanism for light-induced switching to a triplet superconductor by example of two different lattice models, one of which features a metastable chiral topological superconducting state, and discuss implications for real systems such as dilutely-doped 1T' WTe\textsubscript{2}. Our results reveal new guiding principles for engineering metastable unconventional superconducting states using light.

\begin{figure*}[t]
    \centering
    \includegraphics[width=\textwidth]{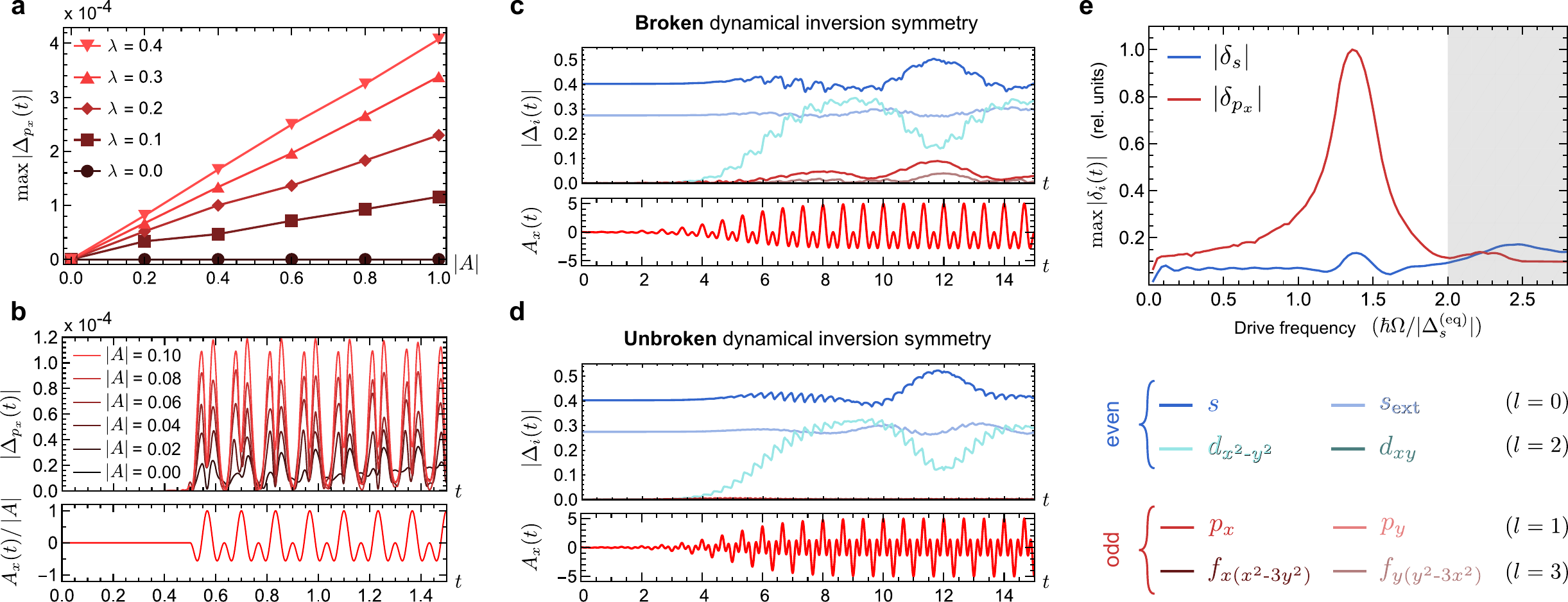}
    \caption{\textbf{Dynamical inversion symmetry breaking and odd-parity Bardasis-Schrieffer (BS) modes.} \textbf{a}-\textbf{b} Demonstration of a linear-in-$|A|$ and linear-in-$\lambda$ (spin-orbit parameter) coupling between an equilibrium order parameter $\Delta_s$ and a $p_x$-wave order parameter $\Delta_{p_x}$. A two-color $x$-polarized pulse that breaks dynamical inversion symmetry is applied with various small magnitudes $|A|$, and example time-plots of the resulting fluctuations in $|\Delta_{p_x}(t)|$ are plotted in \textbf{b}. The maxima of these fluctuations are plotted versus $|A|$ in \textbf{a}, for five different magnitudes of the spin-orbit coupling strength $\lambda$ in units of hopping. \textbf{c} Time dynamics resulting from a two-color pulse that breaks dynamical inversion symmetry (with a $2:1$ frequency ratio). \textbf{d} Time dynamics resulting from a two-color pulse that preserves dynamical inversion symmetry (with a $3:1$ frequency ratio). Both \textbf{c} and \textbf{d} start from purely even superconducting order (indicated by blue curves), and the onset of odd orders (indicated by red and pink curves) is suppressed when dynamical inversion symmetry is unbroken (see legend in bottom right). \textbf{e} Short-time deviations in the $s$-wave and $p_x$-wave order parameters for weak driving with $x$-polarized light as a function of frequency $\Omega$, revealing a BS mode for the $p_x$-wave order parameter that is sub-gap (indicated by the unshaded region).}
    \label{fig:microscopic}
\end{figure*}

\section*{Results}

\subsubsection*{Dynamical inversion symmetry breaking with light}
\label{MicroscopicModel}

While conventional centrosymmetric superconductors typically comprise local spin-singlet Cooper pairs with even parity and net spin $S=0$, spin-triplet superconductivity entails odd-parity pairs with net spin $S=1$ and three allowed values $m_s = 0, \pm 1$ for the spin component. Spin and parity are linked by necessity; since the combination of inversion and spin-exchange has the net effect of exchanging two fermions, even-parity pairs must be singlets while odd-parity pairs must be triplets. Coupling singlet and triplet orders therefore immediately requires the breaking of both inversion and $SU(2)$ spin rotation symmetries. The latter is readily broken in superconductors with strong spin-orbit coupling such as heavy-fermion compounds. Most notably, monolayer $1\text{T}'\,\text{W} \text{Te}_2$ has been observed to host both a quantum spin hall insulating phase \cite{Tang2017,Sajadi2018,Fatemi2018,Wu2018} due to strong spin-orbit coupling and a proximal superconducting phase for dilute doping \cite{Fatemi2018} with a rich array of predicted competing superconducting states with different pairing symmetries \cite{Hsu2020}, rendering it a prime candidate to search for an out-of-equilibrium topological superconducting phase.

In contrast, inversion-breaking turns out to be more subtle, requiring careful consideration of the optical driving scheme. A simple monochromatic light wave is not sufficient, since it preserves a more general dynamical inversion symmetry defined by a combination of parity and time translation by half the wave period $T$,
\begin{equation}
    \mathbf{r} \to -\mathbf{r},\,\,\,\,\,\,\,t\to t-T/2
    \text{.}\label{dyn inv sym}
\end{equation}
This dynamical symmetry can be strongly broken in two ways: (a) via the envelope of an ultrafast pulse, or (b) via a two-color pulse $\mathcal{\mathbf{E}}(t) = \mathcal{E}_1 \cos(\omega_1 t) + \mathcal{E}_2 \cos(\omega_2 t)$ if the two constituent frequencies are not odd harmonics $\omega_1 = p \omega_0,~ \omega_2 = q \omega_0$ of a common frequency $\omega_0$ \cite{Ben-Tal1993, Alon1998, Neufeld2019}. Dynamical symmetries have been extensively studied as a means of high-harmonic generation in atomic and molecular systems \cite{Kfir2015,Baykusheva2016, Neufeld2019_2} and are more recently being used for optically controlling solids \cite{Sederberg2020, Jimenez-Galan2020, Neufeld2021, Trevisan2022}.

We first illustrate the ramifications of breaking these two symmetries for a minimal mean-field model of a centrosymmetric honeycomb lattice superconductor
\begin{align}
    \hat{H} &= -t\sum_{\left<ij\right>\sigma} \CD{i\sigma} \C{j\sigma} - i\lambda \sum_{\left<\left<ij\right>\right>\sigma} \sigma \nu_{ij} \CD{i\sigma} \C{j\sigma}  \notag\\
    &+ U \sum_i \hat{n}_{i\uparrow} \hat{n}_{i\downarrow} + U' \hspace{-0.15cm} \sum_{\left<ij\right>\sigma\sigma'} \hat{n}_{i\sigma} \hat{n}_{j\sigma'}
\end{align}
with effective attractive local ($U$) and nearest-neighbor ($U'$) interactions. Importantly, the inclusion of spin-orbit coupling $\lambda$ via spin-dependent next-nearest-neighbor hopping with phases $\nu_{ij} = \pm 1$ for left or right turns \cite{Kane2005} reduces spin rotation symmetry to $U(1)$, permitting a coupling between singlet and $m_s = 0$ triplet pairs $\sim (\C{-\mathbf{k}\uparrow} \C{\mathbf{k}\downarrow} \mp \C{-\mathbf{k}\downarrow} \C{\mathbf{k}\uparrow} )/\sqrt{2}$.

A standard BCS mean-field decoupling of the interaction in the Cooper channel introduces the superconducting gap function $\Delta_{\alpha\beta}(\mathbf{k}) = \sum_i f^{i}_{\alpha\beta}(\mathbf{k}) \Delta_i$ with sublattice indices $\alpha, \beta$ which can be decomposed into pairing channels
\begin{equation}
    \Delta_i = \frac{v_i}{L^d} \sum_{\mathbf{k}} ~ \overline{f}^i_{\alpha\beta}(\mathbf{k}) \langle \hat{c}_{-\mathbf{k}\beta\downarrow} \hat{c}_{\mathbf{k}\alpha\uparrow} \rangle
\end{equation}
classified in terms of the irreducible representations of the crystal point group with form factors $f^i_{\alpha\beta}(\mathbf{k})$ and channel-projected interactions $v_i$ that depend on $U$, $U'$ [see Methods].

Suppose now that a conventional $s$-wave superconducting phase in equilibrium is irradiated with a weak but wide pump pulse, which couples to electrons via the Peierls substitution with a vector potential $\mathbf{A}(t)$ polarized in the $x$ direction. The pulse is parameterized via the dimensionless field strength $|A| = e a_0 \mathcal{E}_0 / \hbar \omega$ where $\mathcal{E}_0$, $\omega$ and $a_0$ denote the electric field amplitude, frequency, and the lattice constant, respectively. { For numerical expediency, in Fig \ref{fig:microscopic} we use effective interactions $U=U'=-2$ and $\beta=10$, in units of hopping.} Strikingly, the onset of $p_x$-wave order parameter oscillations for weak light pulses is linear in the field strength, shown in Fig. \ref{fig:microscopic}\textbf{a}-\textbf{b} for a two-color pulse, and in stark contrast to $\sim A^2$ scaling for ordinary amplitude mode oscillations. Furthermore, the amplitude of $p_x$ order scales linearly with $\lambda$, completely vanishing in the SU(2) symmetric limit $\lambda=0$. To illustrate the role of inversion symmetry breaking, Fig. \ref{fig:microscopic}\textbf{c} and \textbf{d} depict the order parameter response to a two-color pulse with $2:1$ and $3:1$ frequency ratio, respectively. A $3:1$ frequency ratio preserves a dynamical inversion symmetry [Eq. (\ref{dyn inv sym})] with time translation $t \to t + \pi/\omega$; $p$-wave order oscillations are consequently dramatically suppressed at short times. Conversely, broken dynamical inversion symmetry efficiently excites odd-parity order already for short times [Fig. \ref{fig:microscopic}\textbf{c}]. Note that, due to the nonlinear nature of the quasiparticle equations of motion, these heuristics only apply for time scales of only a few cycles of the pump pulse. Central to efficient switching to triplet order, Fig. \ref{fig:microscopic}\textbf{e} reveals a BS mode resonance for subdominant $p$-wave order parameter oscillations that crucially lies below the ordinary Higgs mode and pair breaking excitations. Spectroscopic observation of this mode would immediately provide an experimentally accessible handle to probe the existence of a subdominant pairing channel, and would importantly suggest that stronger ultrafast excitation of this mode can potentially nudge the system well beyond linear order parameter oscillations and into a metastable competing phase with triplet pairing.

\subsubsection*{Optical switching to a metastable state}
\label{Phenomenological Model}

Insight into whether strongly driving a triplet BS mode can allow for light-induced switching to a metastable odd-parity superconductor can be readily gleaned from an effective time-dependent Ginzburg-Landau (TDGL) description, which encodes the coupling of multiple order parameters to light and importantly accounts for relaxation. In this picture, a suitably tailored pulse liberates the superconducting order parameter from its global free energy minimum and brings it close enough to a proximal local minimum that it relaxes into a metastable opposite-parity phase. A minimal Lagrangian that describes this process reads
\begin{equation}
    \mathcal{L} = -\overline{\Delta}_i\left(\Gamma^{(2)}_{ij}\,\partial_t^2+\Gamma^{(1)}_{ij}\,\partial_t\right)\Delta_j - \beta\mathcal{F}(\Delta_i,\mathbf{\nabla}\Delta_i,\beta)
    \text{.} \label{Lagrangian}
\end{equation}
and includes a kinetic contribution with damping coefficients $\Gamma^{(1)}_{ij}$ and inertial coefficients $\Gamma^{(2)}_{ij}$. The equilibrium free energy $\mathcal{F}$ is taken to generalize the usual Ginzburg-Landau action to $N$ order parameters that crucially include subdominant orders not stabilized in equilibrium, and formally reads
\begin{equation}
\begin{split}
    \beta\mathcal{F} &=~ \mathcal{A}_{ij}\overline{\Delta}_i\Delta_j + \frac{1}{2}\mathcal{B}_{ijmn}\overline{\Delta}_i \Delta_j \overline{\Delta}_m \Delta_n \\
    &+~ \mathcal{C}^\mu_{ij}\overline{\Delta}_i \nabla^\mu \Delta_j + \mathcal{D}^{\mu\nu}_{ij}\nabla^\mu\overline{\Delta}_i\nabla^\nu\Delta_j
    \text{.} \label{FN}
\end{split}
\end{equation}
Here, the bar denotes complex conjugation, and summation over repeated indices is implied. Coefficients $\mathcal{A}_{ij}$, $\mathcal{B}_{ijmn}$, and $\mathcal{D}_{ij}^{\mu\nu}$ are tensorial generalizations of the usual Ginzburg-Landau coefficients for quadratic, quartic, and gradient contributions. We use subscript Latin indices ($i,j,m,n = 1,\hdots,N$) to index different order parameters, and superscript Greek indices ($\mu,\nu$) to index spatial directions. The theory accurately represents a multi-dimensional free energy landscape in the vicinity of the critical temperature $T_c$ for the equilibrium even-parity instability. Finally, we couple the superconductor to light by introducing minimal coupling to a gauge field in velocity gauge,
$\nabla^\mu \longrightarrow \nabla^\mu + i\frac{2e}{\hbar} A^\mu(t)$,
where $-e<0$ is the electron charge. We discard subsequent gradient terms by considering only spatially homogeneous irradiation and order parameters ($\mathbf{\nabla}\Delta_i = 0$). From this, one can derive Euler-Lagrange equations of motion $\left(\frac{\partial\mathcal{L}}{\partial\overline{\Delta}_i} - \partial_t\frac{\partial\mathcal{L}}{\partial(\partial_t\overline{\Delta}_i)} = 0\right)$ that describe light-induced dynamics of competing orders.

The structure of the TDGL action is dictated solely by parity and angular momentum conservation, and importantly permits a coupling between even- and odd-parity order parameters already to linear order in $A^\mu(t)$. If the lattice has $C_n$ rotational symmetry with $n \geq 2$, order parameters for $s$-, $p$-, $d$-wave instabilities can be enumerated by their angular momentum eigenvalues $e^{i2\pi l/n}$, with $l=0$, $l=\pm 1$, $l=\pm 2$ (modulo $n$) respectively. An appealing selection rule permits light-induced coupling $\mathcal{C}_{ij}^\mu$ between superconducting orders $i$, $j$ with $\Delta l = \pm 1$ at linear order in the field. Notably, this dictates that the Lagrangian couples an equilibrium $s$-wave order parameter to odd-parity $p$-wave order already at linear order in $A^\mu(t)$, in agreement with the quasiparticle dynamics of Fig. \ref{fig:microscopic}(a). Conversely, second order in $A^\mu(t)$ contributions couple same-parity order parameters with $\Delta l = 0$ or $\Delta l = \pm2$, capturing the excitation of the conventional amplitude mode. This observation has intriguing consequences for Higgs mode experiments (see Summary and outlook).

\begin{figure}[t]
    \centering
    \includegraphics[width=0.44\textwidth]{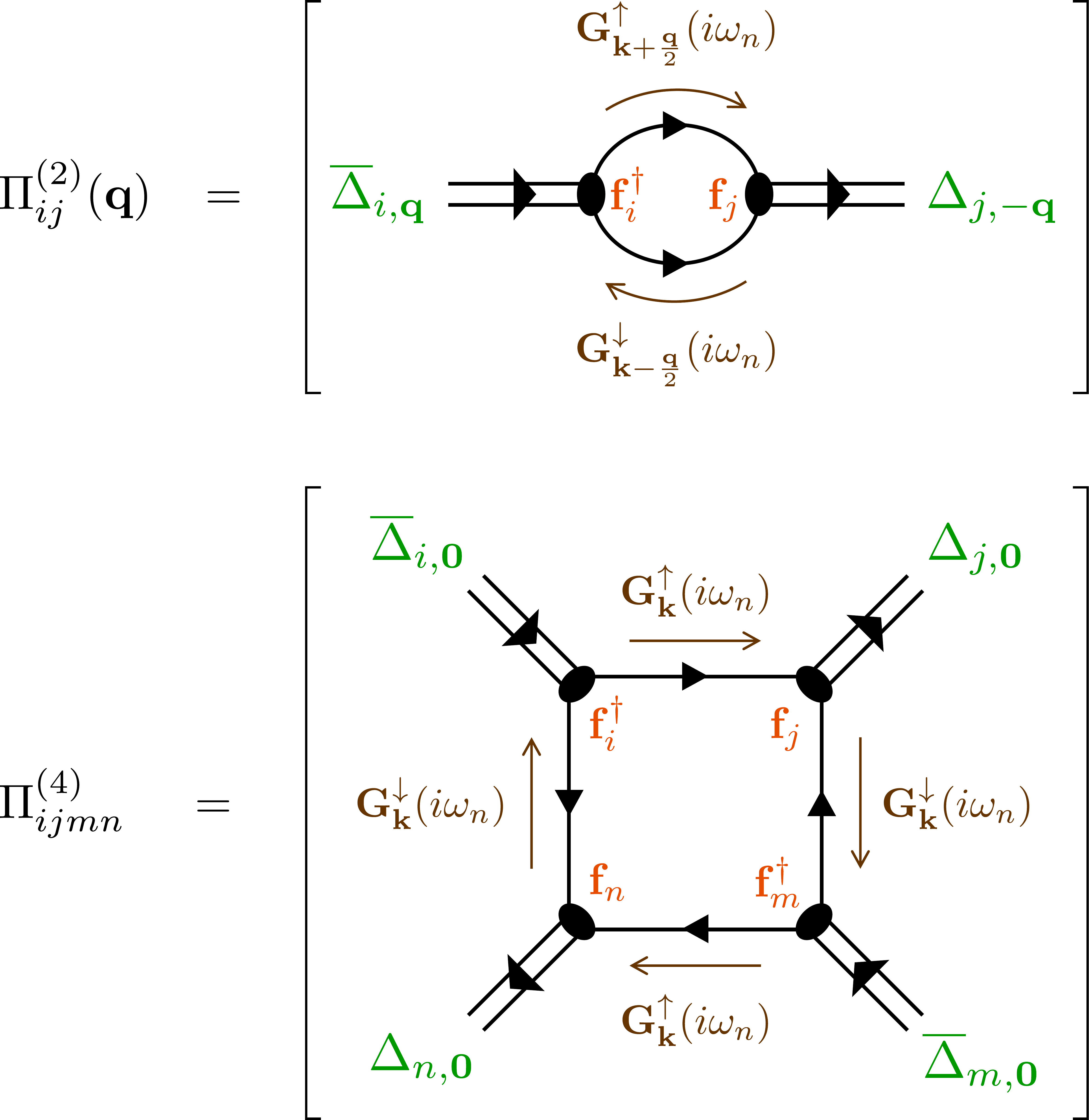}
    \caption{\textbf{Effective Action via Diagrams.} Diagrammatic representation of Eqs. (\ref{Pi2}) and (\ref{Pi4}), which summarize how to calculate the generalized Ginzburg-Landau coefficients as an effective field theory starting from a microscopic Hamiltonian.}
    \label{fig:diagrams}
\end{figure}

To address the existence of a metastable triplet superconductor as a local minimum of the free energy landscape, we explicitly compute the multi-component Ginzburg-Landau coefficients for microscopic Hamiltonians, thus connecting the top-down phenomenological approach to the bottom-up microscopic approach of the previous section. Starting from a generic multiband Hamiltonian with effective attractive pairing interactions, the generalized Ginzburg-Landau coefficients can be computed diagrammatically as
\begin{equation}
    \mathcal{A}_{ij} = -\frac{1}{v_{(i)}}\delta_{ij} + \Pi^{(2)}_{ij}(\mathbf{0}),\,\,\,\,\,\,\,\mathcal{B}_{ijmn} = \frac{1}{2}\Pi^{(4)}_{ijmn}
    \text{,}
\end{equation}
with correlation functions
\begin{align}
    \Pi^{(2)}_{ij}(\mathbf{q}) &= \text{Tr}\left\{\mathbf{f}^\dagger_{i,\mathbf{k}}\mathbf{G}^\uparrow_{\mathbf{k}-\frac{\mathbf{q}}{2}}\mathbf{f}_{j,\mathbf{k}}\mathbf{G}^\downarrow_{\mathbf{k}+\frac{\mathbf{q}}{2}}\right\}
    \text{,}\label{Pi2} \\
    \Pi^{(4)}_{ijmn} &= \text{Tr}\left\{\mathbf{f}^\dagger_{i,\mathbf{k}}\mathbf{G}^\uparrow_{\mathbf{k}}\mathbf{f}_{j,\mathbf{k}}\mathbf{G}^\downarrow_{\mathbf{k}}\mathbf{f}^\dagger_{m,\mathbf{k}}\mathbf{G}^\uparrow_{\mathbf{k}}\mathbf{f}_{n,\mathbf{k}}\mathbf{G}^\downarrow_{\mathbf{k}}\right\}
    \text{,}\label{Pi4}
\end{align}
depicted in Fig. \ref{fig:diagrams}, and Matsubara Green's functions for particles and holes given by
\begin{equation}
    \mathbf{G}^\uparrow_{\mathbf{k}}(i\omega_n) = \frac{1}{i\omega_n-\mathbf{h}_{\mathbf{k}\uparrow}}~,\,\,\,\,\, \mathbf{G}^\downarrow_{\mathbf{k}}(i\omega_n) = \frac{1}{i\omega_n+\mathbf{h}^\top_{-\mathbf{k}\downarrow}}
    ~\text{.}
\end{equation}
In the above equations, $\text{Tr} \equiv \frac{1}{\beta L^d}\sum_{\mathbf{k},\omega_n}\text{tr}$, indicating a sum over momenta, orbital indices, and Matsubara frequencies. The gradient terms $\mathcal{C}^{\mu}_{ij}$ and $\mathcal{D}^{\mu\nu}_{ij}$ are calculated from expanding $\Pi^{(2)}_{ij}(\mathbf{q})$ in powers of $\mathbf{q}$.

\begin{figure*}[t]
    \centering
    \includegraphics[width=\textwidth]{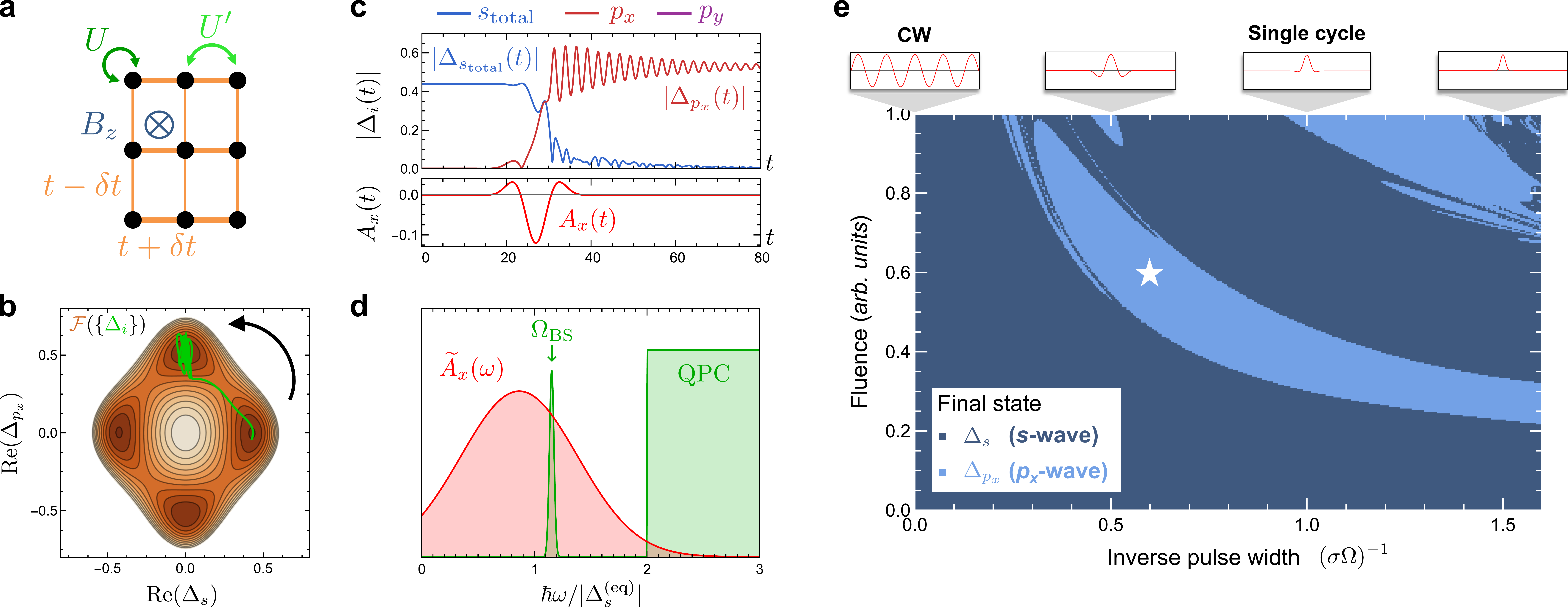}
    \caption{\textbf{Engineering metastable triplet superconductors with light.} \textbf{a} Schematic of a minimal rectangular tight-binding model of competing singlet and triplet pairing instabilities. \textbf{b} Free energy landscape plot of $\text{Re}(\Delta_{p_x})$ vs $\text{Re}(\Delta_s)$, with the time-dependent trajectory in \textbf{c} plotted in green. \textbf{c} Time plots of $A_x(t)$ and $|\Delta_i(t)|$ for ${s_\text{total}}$ [Methods], $p_x$, and $p_y$ order parameters subjected to an example Gaussian pulse. \textbf{d} Schematic plot showing the spectral profile of the pulse in \textbf{c} in relation to the Bardasis-Schrieffer mode frequency $\Omega_\text{BS}$ and the edge of the quasiparticle continuum (QPC). \textbf{e} A parametric plot of fluence versus inverse pulse width showing the final state induced by a family of Gaussian pulses, with frequency $\Omega = 0.75\,\Omega_\text{BS}$ centered slightly below the BS mode frequency. A star indicates the pulse parameters used in \textbf{b}-\textbf{d}.}
    \label{fig:rectangular}
\end{figure*}

We first demonstrate key qualitative features by example of the conceptually simplest single-band model on a rectangular lattice [Fig. \ref{fig:rectangular}\textbf{a}] that captures the essential physics for switching to metastable odd-parity superconductor.
We choose a hopping anisotropy $\delta t = 0.2$, {chemical potential $\mu=-0.1$, and interaction parameters $v_{s} = -1.75$, $v_{p_x}=v_{p_y}=-2.5$, in units of hopping.} As spin-orbit interactions in centrosymmetric crystals require at least a two-orbital unit cell, we break $SU(2)$ via a small Zeeman magnetic field $\Delta_\text{Zeeman}=0.1$ [see Methods]. The free energy parameters are computed from the microscopic model, and the phenomenological inertial/damping coefficients { are set to $\gamma=1.0$, $\eta=0.1$}. The temperature $T = 0.1$ lies below the critical temperature for the $s$-wave and $p_x$-wave channel, with $s$-wave pairing stabilized in equilibrium. Crucially, the coupling between these channels is first-order in $\mathbf{A}(t)$. For simplicity, we assume that pairing interactions in the $d$-wave and extended $s$-wave channels vanish. Furthermore, an $x$-polarized pulse couples solely to $p_x$ order, yielding an appealingly simple minimal action
\begin{equation}
\begin{split}
    \mathcal{L} &= \sum_{i=s,p} \left[ \overline{\Delta}_i \left( \gamma \partial_t^2 + \eta \partial_t \right) \Delta_i + a_i \left|\Delta_i\right|^2 + b_i \left|\Delta_i\right|^4 \right]\\
    &+ \sum_{i=s,p} \left( \nabla^\mu - i \tfrac{2e}{\hbar} A^\mu \right) \overline{\Delta}_i \left( \nabla^\mu + i \tfrac{2e}{\hbar} A^\mu \right) \Delta_i \\
    &+ \frac{1}{2}b_{sp}\left( 4\left| \Delta_{s} \right|^2 \left| \Delta_{p} \right|^2 + \overline{\Delta}_s^2\Delta_p^2 + \overline{\Delta}_p^2\Delta_s^2\right) \\
    &+ \left[ c_{sp} \overline{\Delta}_p \left( \nabla^x + i \tfrac{2e}{\hbar} A^x \right) \Delta_s + \textrm{c.c.} \right] ~\text{,}
    \label{eff action}
\end{split}
\end{equation}
where we abbreviate $\mathcal{A}_{ii} \equiv a_i$, $\mathcal{B}_{iiii} \equiv b_i$, and $\mathcal{B}_{iijj} \equiv b_{ij}$, and assume equal inertial ($\gamma$) and damping ($\eta$) coefficients without loss of generality. The first two lines describe decoupled TDGL actions for $s$ and $p \equiv p_x$ order parameters. The third line describes their quartic coupling that dictates the emergence of a metastable minimum, and the last line couples $s$ and $p$ orders to linear order in the light field $A^\mu$.
{ In the rectangular model, this coupling keeps the $s$- and $p_x$-wave order parameters completely real, allowing the dynamics to be completely captured by a visualizable two-dimensional free energy contour plot} (see Figure \ref{fig:rectangular}\textbf{b}).

Intriguingly, as a function of pulse width and fluence, one finds a contiguous parametric family of Gaussian pulses that result in the final state settling into the pure $\Delta_{p_x}$ stationary point of the free energy, as in Figure \ref{fig:rectangular}\textbf{e}. The presence of a clear threshold for the width of the Gaussian pulse is evidence that dynamical inversion symmetry breaking, which occurs more strongly with a tighter pulse width, is crucial for efficient coupling between opposite-parity orders. We find switching to be most reliable { for strong driving near} the Bardasis-Schrieffer mode frequency [Fig. \ref{fig:microscopic}d], which reads
\begin{equation}
    \Omega_{\text{{BS}}} = \sqrt{\frac{1}{\gamma_{p}}\left(a_{p} - 3 a_{s}\frac{b_{sp}}{b_{s}}\right)}
    \text{,}
\end{equation}
and is determined by deviations from the global free energy minimum in the direction of the target instability [see Methods].
Light polarized in the $x$-direction drives the order parameter strongly in the direction of a closely competing $p_x$ minimum in its free energy landscape, and for Gaussian pulses centered at this frequency with widths on the order of a few resonant periods (see Figure \ref{fig:rectangular}\textbf{e}), the system is efficiently switched to the target metastable state. { Figures 4\textbf{d} and 4\textbf{e} together reveal a  trade-off whereby the pulse needs to be short enough to allow for strong dynamical inversion symmetry breaking while being wide enough (i.e. spectrally sufficiently narrow) to avoid excitation of quasiparticles across the gap (excitations that are not captured by TDGL). The latter constraint leads us to detune the central frequency of the Gaussian pulse to be slightly below $\Omega_\text{BS}$, decreasing the spectral overlap with the quasiparticle continuum while still maintaining a large overlap with the BS mode. [See the Supplementary Material for further details.]}

\begin{figure*}[t]
    \centering
    \includegraphics[width=\textwidth]{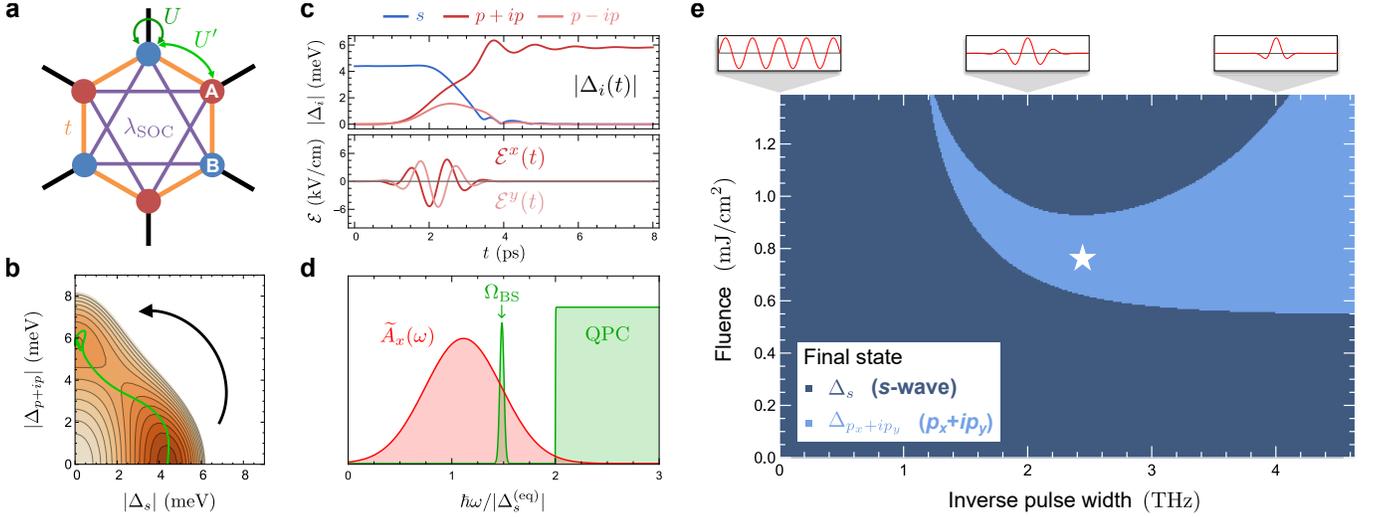}
    \caption{\textbf{Switching between trivial $s$ and topological triplet $p\pm ip$ pairing in honeycomb superconductors.} \textbf{a} Schematic of a minimal honeycomb-lattice conventional $s$-wave superconductor with a competing chiral triplet pairing instability. \textbf{b} Order parameter trajectory in the free energy landscape as a function of the magnitudes of the $s$ and $p+ip$ order parameters, demonstrating the stable switching of the equilibrium order parameter to a metastable chiral state. A small difference in the relaxation rates for the two chiral states is assumed. \textbf{c} Dynamics of $A^x(t)$, $A^y(t)$, and $\left|\Delta_{i}(t)\right|$ for $\Delta_{s}$ and $\Delta_{p\pm i p}$. \textbf{d} Schematic plot of the spectral profile of the pulse in \textbf{c}, showing strong overlap with the BS mode frequency $\Omega_\text{BS}$ and minimal overlap with the quasiparticle continuum (QPC). \textbf{e} Parametric plot in terms of inverse pulse width (in THz) and fluence (in mJ$/$cm$^2$) showing which Gaussian pulses centered at $\Omega=0.75\,\Omega_\text{BS}$ lead to successful switching to the metastable $p+ip$ state. The window is chosen so that no more than $10\%$ of the fluence overlaps with the QPC. A star indicates the pulse parameters used in \textbf{b}-\textbf{d}.}
    \label{fig:honeycomb}
\end{figure*}

We now turn to a realistic model of a conventional centrosymmetric superconductor with strong spin-orbit coupling on the honeycomb lattice, with a putative closely-competing triplet pairing instability [Fig.~\ref{fig:honeycomb}a]. { This models the low-energy physics near the Dirac points of several systems of recent interest, including kagome superconductors \cite{Mielke2021,Shi2022,Chakrabortty2023}, honeycomb materials with large spin-orbit coupling and recently-reported superconductivity \cite{PhysRevLett.111.136804,Liao2018,Zhao2020b}, and many moir\'e heterostructures of transition metal dichalcogenides \cite{Naik2018,Xian2021,Angeli2021,Kennes2021,Zhao2022}.} Rotation symmetry and energetics dictate that the chiral triplet states $\Delta_{p\pm i p} \equiv \frac{1}{\sqrt{2}}\left(\Delta_{p_x}\pm i\Delta_{p_y}\right)$ are stable local free energy minima, while the nodal $\Delta_{p_x}$ and $\Delta_{p_y}$ states individually are not. Figure \ref{fig:honeycomb}\textbf{c} depicts the free energy landscape with an equilibrium $s$-wave minimum and proximal $p \pm ip$ local minima. This is therefore a promising example of a truly metastable non-equilibrium chiral superconducting state in a system with conventional $s$-wave order in equilibrium.

{ The key parameter allowing singlet-triplet switching at first order in the gauge field $\mathbf{A}(t)$ is the coefficient $C^\mu_{ij}$ in Eq. (\ref{FN}). For the honeycomb model, up to first order in the spin-orbit coupling strength $\lambda$, this matrix element
\begin{equation}
\begin{split}
    \mathcal{C}^\mu_{ij} \approx \frac{i\lambda}{2}&\int\frac{d^2\mathbf{k}}{(2\pi)^2}\sum_{mn}\frac{n_\text{F}(\xi_{n\mathbf{k}})-n_\text{F}(-\xi_{m\mathbf{k}})}{\xi_{n\mathbf{k}}+\xi_{m\mathbf{k}}} \\
    & \times \partial_\lambda\left(\left\{\boldsymbol{\mathcal{A}}^\mu_{\mathbf{k}},\mathbf{f}^\dagger_{i,\mathbf{k}}\right\}_{nm}\left(\mathbf{f}_{j,\mathbf{k}}\right)_{mn}\right. \\
    &~~~~~~-\left.\left.\left(\mathbf{f}^\dagger_{i,\mathbf{k}}\right)_{nm}\left\{\boldsymbol{\mathcal{A}}^\mu_{\mathbf{k}},\mathbf{f}_{j,\mathbf{k}}\right\}_{mn}\right)\right|_{\lambda \to 0} + \mathcal{O}(\lambda^2)
    \text{,}
\end{split}
\end{equation}
becomes a function of the (matrix-valued) non-Abelian Berry connection $\mathcal{A}^\mu_{\mathbf{k}mn} \equiv i\left\langle u_{m\mathbf{k}} | \partial_{k_\mu} u_{n\mathbf{k}}\right\rangle$.
Since the energy eigenvalues of the Hamiltonian are at order $\lambda^2$ and higher, the dominant contribution to $\mathcal{C}^\mu$ for small $\lambda$ is therefore purely geometric in nature, controlled by the Berry connection. Notably, this effect is distinct from quantum-geometric corrections to the superfluid weight \cite{peotta2015,torma2022}, and can be equivalently expressed via the Fermi surface Berry connection at low temperatures (see Methods).}

To approximate the energetics in conventional but strongly spin-orbit coupled superconducting systems, \cite{Black-Schaffer2007, Wu2020} we choose parameters so that the nearest neighbor hopping amplitude is $1 \text{ eV}$, the spin-orbit strength is $0.1 \text{ eV}$, the system is hole-doped with chemical potential $\mu = -0.4 \text{ eV}$ and the critical temperatures of the $s$ and $p\pm ip$ states are $\sim 10 \text{ K}$ for effective renormalized interactions $v_s = -1.25 \text{ eV}$, $v_p = -2.3 \text{ eV}$. In this scenario, the superconducting gaps are between $1-10 \text{ meV}$. The inertial coefficients $\Gamma^{(2)}_{ij}$ in Eq. (\ref{Lagrangian}) are set so that the frequency of small amplitude oscillations of $\Delta_i$ matches the expected Higgs mode frequency, $\Omega_i = 2 |\Delta_i^{(\text{eq})}|/\hbar$. This sets the time scale of interest to be on the order of picoseconds, with relevant frequencies between $1-10 \text{ THz}$. The time scale for relaxation (parametrized by $\Gamma^{(1)}_{ij}$ in Eq. (\ref{Lagrangian})) is longer than the Higgs mode period by the dimensionless factor $\beta|\Delta_i^{(\text{eq})}|$ \cite{Yu2021}, which in our case is about $10$. Lastly, assuming a lattice constant on the order of Angstroms, our results call for peak field strengths on the order of $\sim 100 \text{ kV/cm}$ for ultrafast pulses.

Figure \ref{fig:honeycomb}\textbf{e} reveals a broad region of pulse widths ($\sim 0.1$ ps) and fluences ($\sim 1.0$ mJ/cm$^2$) that lead to successful switching from $s$ to a chiral triplet $p+ip$ state via an ultrafast circularly polarized pulse. The resulting order parameter trajectory in a subspace of the free energy landscape is shown in Fig. \ref{fig:honeycomb}\textbf{c} and illustrates key features of the switching protocol, whereby the dynamical inversion symmetry breaking kicks the order parameter in the vicinity of a chiral instability. To allow the system to settle into one of the chiral $p \pm ip$ states within TDGL theory, we set a slight difference in their relaxation rates ($\eta_\pm = \left(1 \pm \delta\right)\eta$ with $\delta = 0.5$). { Though introduced as a phenomenological parameter, one can expect this to arise microscopically from pump-induced time-reversal symmetry breaking of the environmental degrees of freedom that provide dissipation, an effect which is not captured by TDGL theory. The chiral states are} stable against fluctuations in the amplitudes and relative phases of all competing order parameters that are supported by nearest-neighbor pairing interactions, providing a promising proof-of-concept for engineering a metastable topological superconducting state.

\subsubsection*{Summary and outlook}

We present light-induced dynamical inversion symmetry breaking as a generic route to explore competing triplet pairing instabilities in conventional superconductors with strong spin-orbit coupling. Identifying a sub-gap odd-parity BS mode that couples linearly to light, we find that driving this mode with a tailored two-color or ultrafast light pulse can switch the system to a metastable unconventional superconducting state. We illustrate this mechanism for minimal models of superconductors with closely competing pairing instabilities.

Immediate future directions include searching for experimental signatures of a sub-gap opposite-parity Bardasis-Schrieffer mode, in addition to applying this methodology to realistic tight-binding models of candidate materials, including $1\text{T}'\,\text{W} \text{Te}_2$ and moire heterostructures of transition metal dichalcogenides. Following experiments that use pump-probe microscopy to detect the conventional Higgs mode via a linear scaling between pump intensity and the amplitude of resulting gap fluctuations \cite{Shimano2020,Matsunaga2013,Yuzbashyan2006}, a simple optical protocol may be to search for gap fluctuations scaling with the square root of the pump intensity, indicating an amplitude mode coupling linearly to light. Additionally, an intriguing open problem is under what general conditions does a subleading stationary point in the free energy become a local minimum. We have found that the only examples of metastable states are Kramers partners with chiral pairing, i.e. $p_x\pm ip_y$ order parameters in lattices where the $p_x$ and $p_y$ orders are degenerate. 
It would be fruitful to further understand the necessary conditions for a system to support a metastable superconducting instability. Another interesting direction would be to model heating effects from irradiation with an ultrafast pulse, accounting for microscopic mechanisms for heat dissipation not included in our methods. Lastly, throughout this work we use effective renormalized interaction strengths $v_i$ as phenomenological parameters for emergent pairing at low energy scales. Detailed modeling of these parameters in candidate materials, via renormalization group treatments and incorporating RPA effects \cite{Hackner2023}, is an important direction for future work.

\section*{Methods}

\subsubsection*{Gap function equations of motion via quasiparticle dynamics}

Quasiparticle equations of motion for a superconductor coupled to light are derived for a generic BCS-like Hamiltonian
\begin{equation}
\begin{split}
    \hat{H} =&\,\, \sum_{\mathbf{k}\alpha\beta\sigma}  \hat{c}^\dagger_{\mathbf{k}\alpha\sigma} \, h^\sigma_{\alpha\beta}(\mathbf{k}) \, \hat{c}_{\mathbf{k}\beta\sigma} \\ &+\frac{1}{L^d}\sum_{\mathbf{k}\mathbf{k}'\alpha\alpha'\beta\beta'}  V_{\alpha\beta\alpha'\beta'}(\mathbf{k},\mathbf{k}')\,  \hat{c}^\dagger_{\mathbf{k}\alpha\uparrow} \hat{c}^\dagger_{-\mathbf{k}\beta\downarrow} \hat{c}_{-\mathbf{k}'\beta'\downarrow} \hat{c}_{\mathbf{k}'\alpha'\uparrow}
    \text{,}\label{microscopic Hamiltonian}
\end{split}
\end{equation}
where $\hat{c}^\dagger_{\mathbf{k}\alpha\sigma}$ creates an electron in the state with momentum $\mathbf{k}$, orbital index $\alpha \in \{1,\hdots,N\}$, and spin $\sigma \in \{\uparrow,\downarrow\}$. The dependence of the one-body Hamiltonian $h^\sigma_{\alpha\beta}(\mathbf{k})$ on spin encodes spin-orbit coupling that reduces the $SU(2)$ spin rotation symmetry to spin-$z$ conservation. $L$ denotes the linear system size. We decompose the four-fermion interaction vertex $V_{\alpha\beta\alpha'\beta'}(\mathbf{k},\mathbf{k}')$ in terms of a set of orthonormal basis functions $\left\{f^{i}_{\alpha\beta}(\mathbf{k})\right\}$,
\begin{equation}
    V_{\alpha\beta\alpha'\beta'}(\mathbf{k},\mathbf{k}') = \sum_{i} v_i f^i_{\alpha\beta}(\mathbf{k}) \overline{f}^i_{\alpha'\beta'}(\mathbf{k}')
    \text{.}\label{decompose V}
\end{equation}
A standard mean-field decoupling of the interaction in the Cooper channel introduces the superconducting gap function
\begin{equation}
\begin{split}
    \Delta_{\alpha\beta}(\mathbf{k}) &= \frac{1}{L^d} \sum_{\mathbf{k'}\alpha'\beta'} V_{\alpha\beta\alpha'\beta'}(\mathbf{k},\mathbf{k}') \, \langle \hat{c}_{-\mathbf{k}'\beta'\downarrow} \hat{c}_{\mathbf{k}'\alpha'\uparrow} \rangle \\
    &\overset{\text{def}}{=} \sum_{i} \Delta_i \, f^i_{\alpha\beta}(\mathbf{k})
    \text{,}\label{Delta1}
\end{split}
\end{equation}
where the components $\Delta_i$ are classified in terms of the irreducible representations of the crystal point group. The mean-field decomposition of Eq. (\ref{microscopic Hamiltonian}) can be written in Bogoliubov-de Gennes (BdG) form as
\begin{equation}
    \hat{H} = \sum_{\mathbf{k}} \overline{\Psi}_{\mathbf{k}} \, \boldsymbol{\mathcal{H}}^\text{BdG}_{\mathbf{k}}\, \Psi_{\mathbf{k}} - \sum_i \frac{L^d}{v_i}|\Delta_i|^2
    \text{,}
\end{equation}
where
\begin{equation}
    \Psi_{\mathbf{k}} \overset{\text{def}}{=}
    \begin{pmatrix}
    \hat{c}_{\mathbf{k},1,\uparrow} \\
    \vdots \\
    \hat{c}_{\mathbf{k},N,\uparrow} \\
    \hat{c}^\dagger_{-\mathbf{k},1,\downarrow} \\
    \vdots \\
    \hat{c}^\dagger_{-\mathbf{k},N,\downarrow} \\
    \end{pmatrix}
    \text{,}\,\,\,\,\,\,\,\,\,\,
    \boldsymbol{\mathcal{H}}^\text{BdG}_{\mathbf{k}} =
    \begin{pmatrix}
        \mathbf{h}_{\mathbf{k}\uparrow} & \mathbf{\Delta}_{\mathbf{k}} \\
        \mathbf{\Delta}^\dagger_{\mathbf{k}} & -\mathbf{h}^\top_{-\mathbf{k}\downarrow}
    \end{pmatrix}
    \text{.}
\end{equation}
The equal-time Nambu Green's function
\begin{equation}
\begin{split}
    \boldsymbol{\mathcal{G}}^\text{BdG}_{\mathbf{k}} &\overset{\text{def}}{=} \left\langle \overline{\Psi}_{\mathbf{k}} \otimes \Psi_{\mathbf{k}} \right\rangle \\
    &\equiv
    \begin{pmatrix}
        \left[\left\langle \hat{c}^\dagger_{\mathbf{k}\beta\uparrow} \hat{c}_{\mathbf{k}\alpha\uparrow} \right\rangle\right] & \left[\left\langle \hat{c}_{-\mathbf{k}\beta\downarrow} \hat{c}_{\mathbf{k}\alpha\uparrow} \right\rangle\right] \\ \\
        \left[\left\langle \hat{c}^\dagger_{\mathbf{k}\beta\uparrow} \hat{c}^\dagger_{-\mathbf{k}\alpha\downarrow} \right\rangle\right] & \left[\left\langle \hat{c}_{-\mathbf{k}\beta\downarrow} \hat{c}^\dagger_{-\mathbf{k}\alpha\downarrow} \right\rangle\right]
    \end{pmatrix}
    \text{.}
\end{split}
\end{equation}
obeys equations of motion (setting $\hbar=1$)
\begin{equation}
    i\partial_t \boldsymbol{\mathcal{G}}^\text{BdG}_{\mathbf{k}} = \left[ \boldsymbol{\mathcal{H}}^\text{BdG}_{\mathbf{k}}, \boldsymbol{\mathcal{G}}^\text{BdG}_{\mathbf{k}} \right]
    \text{.} \label{G BdG EOM}
\end{equation}
\begin{equation}
    \left.\boldsymbol{\mathcal{G}}^\text{BdG}_{\mathbf{k}}\right\vert_{t\to-\infty} = n_\text{F-D}\left(\boldsymbol{\mathcal{H}}^\text{BdG}_{\mathbf{k}}\right)
    \text{,} \label{G BdG 0}
\end{equation}
where $n_\text{F-D}(\mathbf{X}) \equiv \left(1 + \text{exp}(\beta\mathbf{X})\right)^{-1}$ is the Fermi-Dirac distribution function.
Combined with the instantaneous self-consistency equation for order parameters, this yields equations of motion for the gap function in terms of the Heisenberg dynamics of the Bogoliubov quasiparticles.

\subsubsection*{Symmetry analysis of the Lagrangian}
Suppose we demand that the Lagrangian be invariant under some symmetry transformation on the order parameters $\Delta_i$ and the gauge field $A^{\mu}(t)$,
\begin{equation}
\begin{split}
    \Delta_j &\,\longmapsto\, \mathcal{D}_{jj'}\,\Delta_{j'} \\
    A^\mu &\,\longmapsto\, \mathcal{R}^{\mu\mu'}A^{\mu'}
    \text{,}
\end{split}
\end{equation}
for tensors $\mathcal{D}$ and $\mathcal{R}$. Let $\mathcal{R}$ be orthogonal ($(\mathcal{R}^\top)^{\mu\nu}\mathcal{R}^{\nu\rho} = \delta^{\mu\rho}$) and choose a basis for $\Delta_i$ such that $\mathcal{D}$ is diagonal ($\mathcal{D}_{ij} \equiv d_{(i)}\delta_{ij}$). One can then show that,
\begin{equation}
    \begin{split}
        \mathcal{R}^{\mu\mu'}\mathcal{N}^{\mu'}_{ij} &= \overline{d}_{(i)}d_{(j)} \mathcal{N}^{\mu}_{ij}~\text{,} \\
        \mathcal{R}^{\mu\mu'} \mathcal{M}^{\mu'\nu'}_{ij}(\mathcal{R}^\top)^{\nu'\nu} &= \overline{d}_{(i)}d_{(j)} \mathcal{M}^{\mu\nu}_{ij}
        ~\text{.}
    \end{split}\label{symm constraints}
\end{equation}
$\mathcal{N}^{\mu}_{ij}$ is nonzero if and only if $\overline{d}_{(i)}d_{(j)}$ is an eigenvalue of $\mathcal{R}$, and $\mathcal{M}^{\mu\nu}_{ij}$ is nonzero if $\overline{d}_{(i)}d_{(j)}$ is a product of two (not necessarily distinct) eigenvalues of $\mathcal{R}$. For example, if one considers $C_n$ rotational symmetry on a lattice with $n\geq 2$, the constraints given by Eq. (\ref{symm constraints}) yield the $l$ selection rules discussed in Results.

\subsubsection*{Effective action from the path integral}

We start with the partition function written in terms of an imaginary-time path integral over the Grassmann fields $\overline{\psi}$, $\psi$,
\begin{equation}
    \mathcal{Z} = \int\mathcal{D}[\overline{\psi},\psi] e^{-\left(S_0[\overline{\psi},\psi] + S_\text{int}[\overline{\psi},\psi]\right)}
    \text{,}
\end{equation}
with a free-electron action given by,
\begin{equation}
    S_0[\overline{\psi},\psi] = \int_0^\beta d\tau \sum_{\mathbf{k}\sigma\alpha\beta} \overline{\psi}_{\mathbf{k}\alpha\sigma}\left[\delta_{\alpha\beta}\,\partial_\tau + h^\sigma_{\alpha\beta}(\mathbf{k})\right] \psi_{\mathbf{k}\beta\sigma}
    \text{,}
\end{equation}
and an interacting action given by,
\begin{equation}
\begin{split}
    S_\text{int}[\overline{\psi},\psi] = \int_0^\beta d\tau \, \frac{1}{L^d}\sum_{\mathbf{k}\mathbf{k}'\mathbf{q}\,\alpha\alpha'\beta\beta' i} v_i f^i_{\alpha\beta}(\mathbf{k})\overline{f}^i_{\alpha'\beta'}(\mathbf{k}') \\
    \times\,\overline{\psi}_{\mathbf{k}+\frac{\mathbf{q}}{2},\alpha\uparrow}\overline{\psi}_{-\mathbf{k}+\frac{\mathbf{q}}{2},\beta\downarrow}\psi_{-\mathbf{k}'+\frac{\mathbf{q}}{2},\beta'\downarrow}\psi_{\mathbf{k}'+\frac{\mathbf{q}}{2},\alpha'\uparrow}
    \text{.}
\end{split}
\end{equation}
As before, $\alpha,\beta$ denote orbital indices, $\sigma$ denotes a spin index, and $v_i$ denotes the interaction decomposed into pairing channels. A Hubbard-Stratonovich transformation \cite{Altland2010} decouples $S_\text{int}$ in the Cooper channel in terms of auxiliary fields $\overline{\Delta}_{i,\mathbf{q}}$, $\Delta_{i,-\mathbf{q}}$,
\begin{equation}
\begin{split}
    &\tilde{S}[\overline{\Psi},\Psi,\overline{\Delta},\Delta]\\
    &\,\,\,\,= \sum_{\mathbf{k}\,\mathbf{k}'\,\omega_n} \overline{\Psi}_{\mathbf{k},\omega_n} \left(\boldsymbol{\mathcal{G}}^{-1}\right)^{\omega_n}_{\mathbf{k},\mathbf{k}'} \Psi_{\mathbf{k}',\omega_n} - \sum_{i\,\mathbf{q}} \overline{\Delta}_{i,\mathbf{q}}\frac{\beta L^d}{v_i}\Delta_{i,-\mathbf{q}}
    \text{,}\label{Z matsubara}
\end{split}
\end{equation}
where the Nambu spinor $\Psi$ and Gor'kov Green's operator $\mathbf{\boldsymbol{\mathcal{G}}}^{-1}$ are defined in terms of their momentum components and Matsubara frequency dependence as follows,
\begin{equation}
    \Psi_{\mathbf{k}} \overset{\text{def}}{=}
    \begin{pmatrix}
        \psi_{\mathbf{k},1,\uparrow} \\
        \vdots \\
        \psi_{\mathbf{k},N,\uparrow} \\
        \overline{\psi}_{-\mathbf{k},1,\downarrow} \\
        \vdots \\
        \overline{\psi}_{-\mathbf{k},N,\downarrow} \\        
    \end{pmatrix}
    \text{,}
\end{equation}
\begin{equation}
\begin{split}
&\left(\boldsymbol{\mathcal{G}}^{-1}\right)^{\omega_n}_{\mathbf{k}+\frac{\mathbf{q}}{2},\mathbf{k}-\frac{\mathbf{q}}{2}} \\
&=
    \begin{pmatrix}
        \left(-i\omega_n + \hat{h}_{\mathbf{k}\uparrow}\right)\delta_{\mathbf{q},\mathbf{0}} & \displaystyle \sum_i \Delta_{i,-\mathbf{q}}\, \hat{f}_{\mathbf{k},i} \\
        \displaystyle \sum_i \bar{\Delta}_{i,-\mathbf{q}}\,\hat{f}^\dagger_{\mathbf{k},i} & \left(-i\omega_n - \hat{h}^\top_{-\mathbf{k}\downarrow}\right)\delta_{\mathbf{q},\mathbf{0}}
    \end{pmatrix}
    \text{.}
\end{split}
\end{equation}
Note that here, $\mathbf{k}$ can be interpreted as the internal momentum of a Cooper pair, while $\mathbf{q}$ corresponds to the external momentum. One can then perform the path integral over the Grassmann fields $\overline{\Psi}$, $\Psi$, which gives,
\begin{equation}
    \mathcal{Z} = \int\mathcal{D}[\overline{\Delta},\Delta] \exp\left\{\ln\det \boldsymbol{\mathcal{G}}^{-1} + \sum_{i\,\mathbf{q}} \overline{\Delta}_{i,\mathbf{q}}\frac{\beta L^d}{v_i}\Delta_{i,-\mathbf{q}}\right\}
    \text{.}
\end{equation}
Using $\ln \det \mathbf{M} = \text{tr} \ln \mathbf{M}$, this results in an effective action
\begin{equation}
    S_\text{eff}[\overline{\Delta},\Delta] = -\text{tr} \ln \boldsymbol{\mathcal{G}}^{-1} - \sum_{i\,\mathbf{q}} \overline{\Delta}_{i,\mathbf{q}}\frac{\beta L^d}{v_i}\Delta_{i,-\mathbf{q}}
    \text{.}
\end{equation}
Expanding this to fourth order in $\Delta$ yields,
\begin{equation}
\begin{split}
    S_\text{eff}[\overline{\Delta},\Delta] =&\,\, \beta L^d \sum_{i,j,\mathbf{q}}\overline{\Delta}_{i,\mathbf{q}}\left(-\frac{\delta_{ij}}{v_i}+\Pi^{(2)}_{ij}(\mathbf{q})\right)\Delta_{j,-\mathbf{q}} \\
    &\,\,+\beta L^d \frac{1}{4}\sum_{i,j,m,n} \overline{\Delta}_{i,\mathbf{0}}\overline{\Delta}_{m,\mathbf{0}}\Pi^{(4)}_{ijmn}(\mathbf{0})\Delta_{j,\mathbf{0}}\Delta_{n,\mathbf{0}}
    \text{,}\label{expanded action}
\end{split}
\end{equation}
which, comparing to Eq. (\ref{FN}), allows one to compute the generalized Ginzburg-Landau coefficients for completing orders as described in Results.

\subsubsection*{Model details}
\subsubsection*{Rectangular lattice model}

Consider a rectangular lattice with hopping amplitudes $t_x > t_y > 0$, on-site interaction $U$, and nearest neighbor interaction $U'$. As discussed in the main text, breaking $SU(2)$ symmetry in a single-band model requires including a small the Zeeman splitting due to a $z$-aligned magnetic field. The single-particle dispersion reads
\begin{equation}
    \epsilon_{\mathbf{k}\sigma} = -2(t_x \cos{k_x} + t_y \cos{k_y}) + \Delta_\text{Zeeman}\text{sgn}(\sigma)
    \text{.}
\end{equation}
When the Zeeman splitting energy $\Delta_\text{Zeeman}$ is nonzero, there are separate spin-up and spin-down Fermi surfaces, making it difficult to define singlet/even order parameters and triplet/odd order parameters in this model. For this reason, we will consider the applied magnetic field to be zero initially, only to be adiabatically turned on before the optical pulse and adiabatically turned off after the optical pulse. We note that a small $\Delta_\text{Zeeman}$ has a negligible effect on the Ginzburg-Landau free energy, but must necessarily be present to allow coupling between singlet and triplet orders. Conversely, multiband models discussed below allow for the inclusion of spin-orbit coupling, obviating a magnetic field.

The form factors $f^i(\mathbf{k})$ for this model up to nearest-neighbor interactions are tabulated in Table \ref{form factors rectangular}. { There are three order parameters in the $A$ irrep ($s$-wave) and two in the $B$ irrep ($p$-wave). In equilibrium, the system settles into a combination of the three $s$-wave orders dictated by the relative strengths of the on-site and nearest neighbor interactions. For simplicity, we re-express these order parameters in the eigenbasis of $\Pi_{ij}^{(2)}(\mathbf{0})$ [Eq. (\ref{Pi2})], defining $\Delta_{s}$ in the main text to be the order parameter with the lowest eigenvalue. We then define $|\Delta_{s_\text{total}}|$ in Figure 4\textbf{c} to be $\sqrt{\Delta_{s_\text{local}}^2+\Delta_{s_{\text{ext},x}}^2+\Delta_{s_{\text{ext},y}}^2}$.}

\subsubsection*{Honeycomb lattice model with spin-orbit coupling}

We now consider a honeycomb lattice with Kane-Mele spin-orbit coupling \cite{Kane2005} as well as effective on-site and nearest-neighbor attractive interactions,
\begin{equation}
    \left[h^\sigma_{\alpha\beta}(\mathbf{k})\right] =
    \begin{pmatrix}
        \lambda v_\text{SO}^\sigma(\mathbf{k}) - \mu & g(\mathbf{k}) \\
        g^*(\mathbf{k}) & -\lambda v_\text{SO}^\sigma(\mathbf{k}) - \mu
    \end{pmatrix}
    \text{,}
\end{equation}
with nearest and next-nearest-neighbor hopping
\begin{equation}
    g(\mathbf{k}) = -t \sum_{\mathbf{d}_i} e^{-i\mathbf{k}\cdot\mathbf{d}_i}
    \text{,}~~~ 
    v^\sigma_\text{SO}(\mathbf{k}) = \text{sgn}(\sigma) \sum_{\mathbf{a}_i} \sin(\mathbf{k}\cdot\mathbf{a}_i)
    \text{.}
\end{equation}
Here, $\lambda$ parameterizes the strength of spin-orbit coupling, $\mu$ denotes the chemical potential, $t$ denotes the hopping parameter, $\{\mathbf{d}_i\}$ denotes the three nearest-neighbor lattice vectors, and $\{\mathbf{a}_i\}$ denotes the three next-nearest-neighbor lattice vectors ($\mathbf{a}_1 = \mathbf{d}_2-\mathbf{d}_3$, $\mathbf{a}_2 = \mathbf{d}_3 - \mathbf{d}_1$, $\mathbf{a}_3 = \mathbf{d}_1-\mathbf{d}_2$). For the interaction term, we consider an on-site attraction $U$ and nearest-neighbor attraction $U'$, decomposing $V_{\alpha\beta\alpha'\beta'}(\mathbf{k},\mathbf{k}')$ as in Eq. (\ref{decompose V}) using the basis functions tabulated in Table \ref{form factors honeycomb}. Setting $U=U'$, the $\Delta_{s}$ state is found to have the highest critical temperature, followed by two degenerate $p_x$ and $p_y$ states. These have a $d$-wave $\mathbf{k}$-space structure, but are odd under sublattice exchange and have angular momentum $l=\pm 1$. Chiral superpositions are abbreviated as $\frac{1}{\sqrt{2}}\left(\Delta_{p_x} \pm i \Delta_{p_y}\right) \equiv \Delta_{p \pm i p}$.

\subsubsection*{BS mode frequency and free energy barrier between two order parameters}

The Lagrangian [Eq. (\ref{Lagrangian})] results in Euler-Lagrange equations of motion
\begin{equation}
\begin{split}
    \left(-\Gamma^{(2)}_{ij}\,\partial_t^2 - \Gamma^{(1)}_{ij}\,\partial_t\right)\Delta_j &=
    \left[\mathcal{A}_{ij} + \mathcal{B}_{ijmn}\overline{\Delta}_m\Delta_n\right. \\ &\,\,\,\,\,\,\,\,\,+ \mathcal{N}^\mu_{ij}A^\mu(t) \\
    &\,\,\,\,\,\,\,\,\,\left.+ \mathcal{M}^{\mu\nu}_{ij}A^\mu(t)A^\nu(t)\right]\Delta_j
    \text{,} \label{EOM}
\end{split}
\end{equation}
where $\mathcal{N}^\mu_{ij} \equiv i\frac{2e}{\hbar}\mathcal{C}^\mu_{ij}$ and $\mathcal{M}^{\mu\nu}_{ij} \equiv (i\frac{2e}{\hbar})^2 \mathcal{D}^{\mu\nu}_{ij}$ are defined to absorb factors of $i\frac{2e}{\hbar}$ from minimally substituting $\nabla^\mu \longrightarrow \nabla^\mu + i\frac{2e}{\hbar} A^\mu(t)$ into Eq. (\ref{FN}) and assuming spatially-homogeneous order parameters and irradiation.

Focusing on the simplest case of two competing order parameters that are only coupled at linear order in $A^\mu(t)$ (e.g. $s$-wave and $p_x$-wave), the free energy for two order parameters can be written concisely as,
\begin{equation}
\begin{split}
    \beta\mathcal{F}^{(2)} =&\,\, a_1|\Delta_1|^2+\frac{1}{2}b_1|\Delta_1|^4 + a_2|\Delta_2|^2 + \frac{1}{2}b_2|\Delta_2|^4 \\
    &\,\,+\frac{1}{2}b_{12}\left(4|\Delta_1|^2|\Delta_2|^2 + \overline{\Delta}_1^2\Delta_2^2+\overline{\Delta}_2^2\Delta_1^2\right)
    \text{,}\label{FE2}
\end{split}
\end{equation}
where $a_i\equiv\mathcal{A}_{ii}$, $b_i\equiv \mathcal{B}_{iiii}$, and $b_{12}$ equals any fourth-order coefficient coupling two $\Delta_1$'s and two $\Delta_2$'s (these are all equivalent in this case). This is in agreement with Ref. \cite{Sergienko2004}. Assuming the system is initially in the equilibrium state, $\Delta^{(\text{eq})}_1=\sqrt{-a_1/b_1}, \Delta^{(\text{eq})}_2=0$ (assume overall phase equals zero without loss of generality), the equation of motion for $\Delta_{2}$ up to linear order in $\Delta_{2}$ and $A^\mu(t)$ is given by,
\begin{equation}
\begin{split}
    (-\gamma_2\partial_t^2-\eta_2\partial_t) \Delta_{2}(t) = M^2\Delta_{2}(t) +A^\mu(t)\mathcal{N}^\mu_{21}\Delta^{(\text{eq})}_{1} \\
    + \,\mathcal{O}(|\mathbf{A}(t)|^2,|\Delta_{2}(t)|^3)
    \text{,}
\end{split}
\end{equation}
where $\gamma_2\equiv\Gamma^{(2)}_{22}$, $\eta_2\equiv\Gamma^{(1)}_{22}$  (we assume the inertial and damping coefficients are diagonal), and $M^2$ is given by,
\begin{equation}
    M^2 =  a_2 - 3a_1\frac{b_{12}}{b_1} \equiv \gamma_2 \Omega^2
    \text{.}
\end{equation}
One can formally integrate this equation of motion as,
\begin{equation}
    \Delta_2(t) = \mathcal{N}^\mu_{21}\Delta^{(\text{eq})}_1\int\frac{d\omega}{2\pi}\frac{e^{i\omega t}}{\gamma_2\omega^2-M^2-i\eta_2\omega}\widetilde{A}^\mu(\omega)
    \text{,}
\end{equation}
which yields a resonant response when $\omega^2 \equiv \Omega^2 = M^2/\gamma_2$. { We identify this as the BS mode frequency $\Omega_\text{BS}$. One can also derive from Eq. (\ref{FE2}) the free energy barrier between the two states by finding the saddle point surrounded by the $\Delta_1$ minimum, the $\Delta_2$ minimum, and the $\Delta_{1,2}=0$ maximum. The free energy difference between this saddle point and the global minimum is given by,
\begin{equation}
    E_\text{barrier} = \frac{a_1^2 b_2 + a_2^2 b_1 - 2a_1 a_2 b_{12}(2+\cos{\phi})}{(9+8\cos{\phi}+\cos{2\phi})b_{12}^2-2 b_1 b_2}+\frac{a_1^2}{2b_1}
    \text{,}
\end{equation}
where $\phi \equiv \text{arg}\left(\Delta_2/\Delta_1\right)$ is the relative phase between the two order parameters. For successful switching, the time-integrated pulse must supply enough kinetic energy $\sim\frac{1}{2}\gamma_2|\partial_t\Delta_2|^2$ to overcome this free energy barrier. Since the matrix element $\mathcal{N}^\mu_{ij}\propto\mathcal{C}^\mu_{ij}$ depends linearly on the spin-orbit coupling strength $\lambda$ (for small $\lambda$), we expect the requisite fluence for switching to scale as $\sim (t/\lambda)^2$ (where $t$ is the nearest-neighbor hopping amplitude).

\subsubsection*{$\mathcal{C}^{\mu}_{ij}$ in terms of the Fermi surface Berry connection}

Assuming the chemical potential exists in a band with Bloch functions $|u_\mathbf{k}\rangle$, and dispersion $\xi_\mathbf{k}$ (measured with respect to the chemical potential) with a large gap $\Delta_\text{gap}$ to all other bands that we formally take to infinity, 
\begin{equation}
    \mathcal{C}^\mu_{ij} = \frac{i\lambda}{2}\int\frac{d^2\mathbf{k}}{(2\pi)^2}\frac{\tanh{\frac{\beta\xi_{\mathbf{k}}}{2}}}{2\xi_{\mathbf{k}}}\left.\partial_\lambda \Lambda^\mu_{\mathbf{k},ij}\right|_{\lambda\to 0}+ \mathcal{O}\left(\lambda^2, \Delta_\text{gap}^{-1}\right)
\end{equation}
\begin{equation}
\begin{split}
    \Lambda^\mu_{\mathbf{k},ij} =&~ \big\langle u_\mathbf{k}\big|\mathbf{f}^\dagger_i\big|u_\mathbf{k}\big\rangle\bigg(\big\langle\partial^\mu u_\mathbf{k}|\mathbf{f}_j|u_\mathbf{k}\big\rangle-\big\langle u_\mathbf{k}|\mathbf{f}_j|\partial^\mu u_\mathbf{k}\big\rangle\bigg)\\
    &-\bigg(\big\langle\partial^\mu u_\mathbf{k}|\mathbf{f}^\dagger_i|u_\mathbf{k}\big\rangle-\big\langle u_\mathbf{k}|\mathbf{f}^\dagger_i|\partial^\mu u_\mathbf{k}\big\rangle\bigg)\big\langle u_\mathbf{k}\big|\mathbf{f}_j\big|u_\mathbf{k}\big\rangle
\end{split}
\end{equation}
In the $\beta\to\infty$ limit, the integrand diverges at the Fermi surface, meaning the dominant contribution to $\mathcal{C}^\mu_{ij}$ at low temperatures and low spin-orbit strengths comes from the geometry of the Bloch states at the Fermi surface.}

\section*{Data availability}
Data sets are available from the corresponding author on request.

\section*{Code availability}
Codes are available from the corresponding author on request.

\bibliography{references}

\begin{acknowledgements}
We thank Gene Mele and Dante Kennes for helpful discussions. S.G. is supported by the NSF Graduate Research Fellowship Program under Grant No. DGE-1845298. C.S.W. is supported by the Deutsche Forschungsgemeinschaft (DFG, German Research Foundation) via RTG 1995. M.C. acknowledges support from the NSF under Grant No. DMR-2132591.
\end{acknowledgements}

\section*{Author Contributions Statement}
S.G. developed the theoretical description and performed numerical calculations for the effective action approach. S.G. and C.S.W. performed numerical calculations for the microscopic approach. All authors contributed to writing the manuscript. M.C. designed and supervised the project.

\section*{Competing interests Statement}
The authors declare no competing interests.

\section*{Tables}

\begin{table}[H]
    \centering
    \begin{tabular}{c|c|c}
    \text{Irrep.} & \text{Pairing} & $f^i(\mathbf{k})$ \\
    \hline
    $A$ & $s_\text{local}$ & $1$ \\
    $B$ & $p_x$ & $\sqrt{2}\sin{k_x}$ \\
    $B$ & $p_y$ & $\sqrt{2}\sin{k_y}$ \\
    $A$ & $s_{\text{ext},x}$ & $\sqrt{2}\cos{k_x}$ \\
    $A$ & $s_{\text{ext},y}$ & $\sqrt{2}\cos{k_y}$ \\
    \end{tabular}
    \caption{Basis functions $f^{i}(\mathbf{k})$ that form a complete basis for $V(\mathbf{k},\mathbf{k}') = \sum_i v_i f^i(\mathbf{k})\overline{f}^i(\mathbf{k'})$ up to nearest-neighbor pairing interactions on the rectangular lattice. Normalization is such that $\displaystyle\int\frac{d^2k}{(2\pi)^2} \overline{f}^i(\mathbf{k})f^j(\mathbf{k}) = \delta_{ij}$.}
    \label{form factors rectangular}
\end{table}

\begin{table}[H]
    \centering
    \begin{tabular}{c|c|c}
    \text{Irrep.} & \text{Pairing} & $f^i_{\alpha\beta}(\mathbf{k})$ \\
    \hline
    $A_{1g}$ & $s$ & $\frac{1}{\sqrt{2}}
    \begin{pmatrix}
        1 & 0 \\
        0 & 1
    \end{pmatrix}$ \\
    $A_{1g}$ & $s_\text{ext}$ & $\frac{1}{\sqrt{2}}
    \begin{pmatrix}
        0 & f_{s\text{-ext}}(\mathbf{k}) \\
        \overline{f}_{s\text{-ext}}(\mathbf{k}) & 0
    \end{pmatrix}$ \\
    $E_{1u}$ & $p_x$ & $\frac{1}{\sqrt{2}}
    \begin{pmatrix}
        0 & f_{d_{x^2\text{-}y^2}}(\mathbf{k}) \\
        -\overline{f}_{d_{x^2\text{-}y^2}}(\mathbf{k}) & 0
    \end{pmatrix}$ \\
    $E_{1u}$ & $p_y$ & $\frac{1}{\sqrt{2}}
    \begin{pmatrix}
        0 & f_{d_{xy}}(\mathbf{k}) \\
        -\overline{f}_{d_{xy}}(\mathbf{k}) & 0
    \end{pmatrix}$ \\
    $E_{1g}$ & $d_{x^2\text{-}y^2}$ & $\frac{1}{\sqrt{2}}
    \begin{pmatrix}
        0 & f_{d_{x^2\text{-}y^2}}(\mathbf{k}) \\
        \overline{f}_{d_{x^2\text{-}y^2}}(\mathbf{k}) & 0
    \end{pmatrix}$ \\
    $E_{1g}$ & $d_{xy}$ & $\frac{1}{\sqrt{2}}
    \begin{pmatrix}
        0 & f_{d_{xy}}(\mathbf{k}) \\
        \overline{f}_{d_{xy}}(\mathbf{k}) & 0
    \end{pmatrix}$ \\
    $B_{1u}$ & $f_1$ & $\frac{1}{\sqrt{2}}
    \begin{pmatrix}
        1 & 0 \\
        0 & -1
    \end{pmatrix}$ \\
    $B_{1u}$ & $f_2$ & $\frac{1}{\sqrt{2}}
    \begin{pmatrix}
        0 & f_{s\text{-ext}}(\mathbf{k}) \\
        -\overline{f}_{s\text{-ext}}(\mathbf{k}) & 0
    \end{pmatrix}$
    \end{tabular}
    \\
    \vspace{0.2in}
    \begin{tabular}{c}
    $f_{s\text{-ext}}(\mathbf{k})=\frac{1}{\sqrt{3}}e^{-i\mathbf{k}\cdot\mathbf{d}_1}+\frac{1}{\sqrt{3}}e^{-i\mathbf{k}\cdot\mathbf{d}_2}+\frac{1}{\sqrt{3}}e^{-i\mathbf{k}\cdot\mathbf{d}_3}$ \\
    $f_{d_{x^2\text{-}y^2}}(\mathbf{k})=\sqrt{\frac{2}{3}}e^{-i\mathbf{k}\cdot\mathbf{d}_1}-\frac{1}{\sqrt{6}}e^{-i\mathbf{k}\cdot\mathbf{d}_2}-\frac{1}{\sqrt{6}}e^{-i\mathbf{k}\cdot\mathbf{d}_3}$ \\
    $f_{d_{xy}}(\mathbf{k})=-\frac{1}{\sqrt{2}}e^{-i\mathbf{k}\cdot\mathbf{d}_2}+\frac{1}{\sqrt{2}}e^{-i\mathbf{k}\cdot\mathbf{d}_3}$
    \end{tabular}
    \\
    \vspace{0in}
    \caption{Basis functions $f^{i}_{\alpha\beta}$ that form a complete basis for $V_{\alpha\beta\alpha'\beta'}(\mathbf{k},\mathbf{k}') = \sum_i v_i f^i_{\alpha\beta}(\mathbf{k})\overline{f}^i_{\alpha'\beta'}(\mathbf{k'})$ up to nearest-neighbor pairing interactions on the honeycomb lattice.}
    \label{form factors honeycomb}
\end{table}

\end{document}


\title{Supplementary Material: Light-induced switching between singlet and triplet superconducting states}
\author{Steven Gassner, $\,$ Clara S. Weber, $\,$ Martin Claassen}
\date{}

\maketitle

\section{Quasiparticle excitations and pulse fluence}

In Figures 4 and 5 of the main text, we report parameter regions as a function of inverse pulse width and fluence that lead to successful switching in our time-dependent Ginzburg-Landau (TDGL) simulations. In Figure 5\textbf{e}, we comment that the window of pulse widths was chosen such that fraction of fluence with frequencies in the quasiparticle continuum (defined as $\omega > 2\Delta^{(\text{eq})}$, where $2\Delta^{(\text{eq})}$ is the equilibrium superconducting gap) never exceeds $10\%$. Here we provide plots of this fluence fraction as a function of the pulse parameters. We compute this via the ratio $\int_{2\Delta^{(\text{eq})}}^{\infty} |\mathbf{E}(\omega)|^2 d\omega / \int_{-\infty}^{\infty} |\mathbf{E}(\omega)|^2 d\omega$, with $\mathbf{E}(\omega) = \omega\mathbf{A}(\omega)$. The results are plotted in Figure 1.

\begin{figure}[h]
    \centering
    \includegraphics[width=\textwidth]{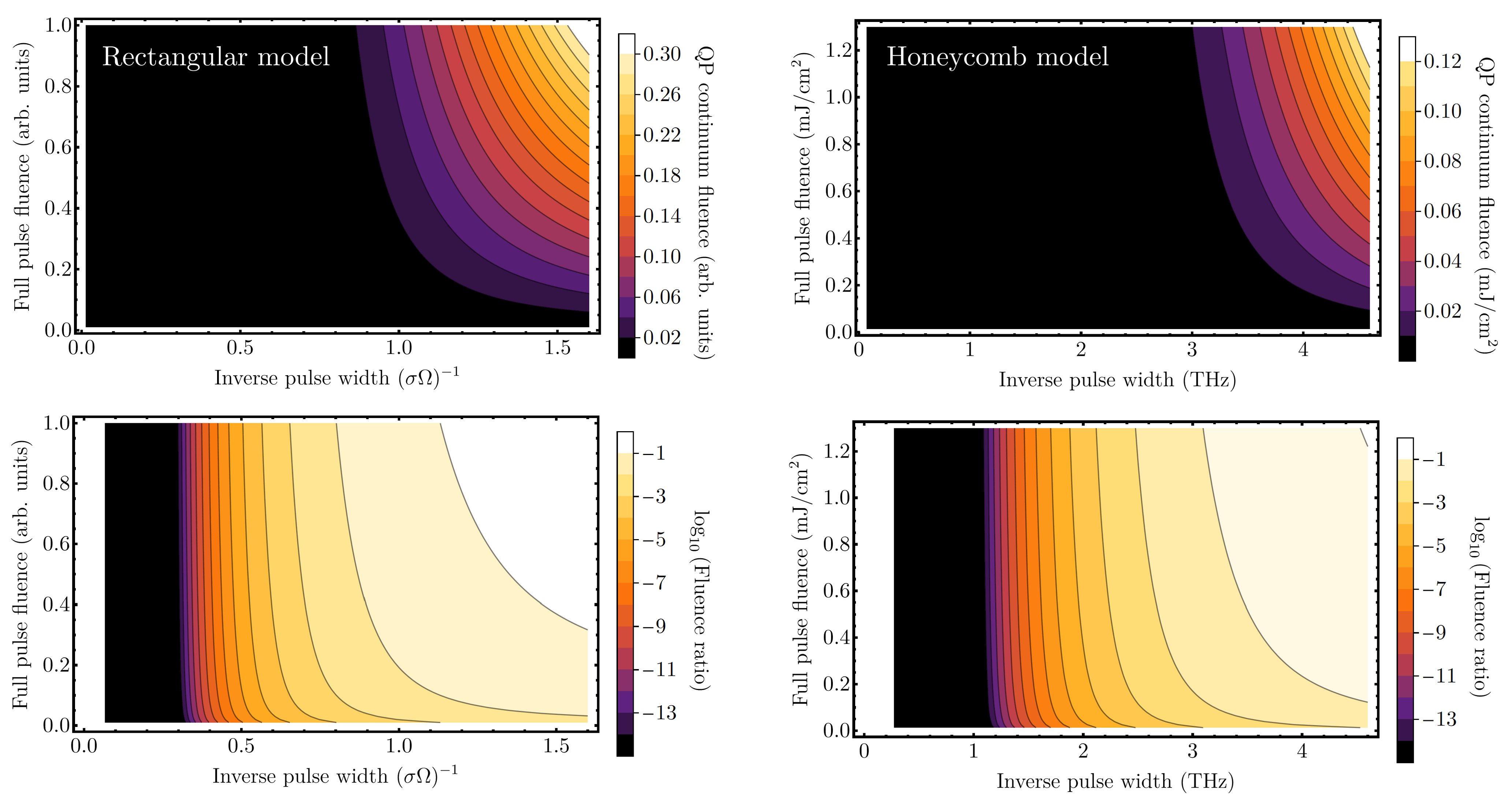}
    \caption{The amount of pulse fluence contributing to quasiparticle excitations as a function of the pulse parameters used in Figure 4 (Left, Rectangular model) and Figure 5 (Right, Honeycomb model). The upper plots calculate the fluence. The lower plots calculate the log of the ratio between the quasiparticle continuum fluence and the full fluence of the pulse.}
    \label{fig:supplement}
\end{figure}

\section{Path integral derivation of the generalized Ginzburg-Landau coefficients}

Here we use a path integral to derive the Ginzburg-Landau effective action presented in the main text from an arbitrary multiband Bloch Hamiltonian. We start with the partition function for the microscopic model written in terms of a path integral over the electronic Grassmann fields $\bar{\psi}$, $\psi$
\begin{equation}
    \mathcal{Z} = \int\mathcal{D}[\bar{\psi},\psi] e^{-\left(S_0[\bar{\psi},\psi] + S_\text{int}[\bar{\psi},\psi]\right)}
    \text{,}
\end{equation}
with a free-electron action given by,
\begin{equation}
    S_0[\bar{\psi},\psi] = \int_0^\beta d\tau \sum_{\mathbf{k}\sigma\alpha\beta} \bar{\psi}_{\mathbf{k}\alpha\sigma}\left[\delta_{\alpha\beta}\,\partial_\tau + h^\sigma_{\alpha\beta}(\mathbf{k}+e\mathbf{A})\right] \psi_{\mathbf{k}\beta\sigma}
    \text{,}
\end{equation}
and an interaction given by,
\begin{equation}
    S_\text{int}[\bar{\psi},\psi] = \int_0^\beta d\tau \, \frac{1}{L^d}\sum_{\mathbf{k}\mathbf{k}'\alpha\alpha'\beta\beta'} V_{\alpha\beta\alpha'\beta'}(\mathbf{k},\mathbf{k}')\bar{\psi}_{\mathbf{k}-\frac{\mathbf{q}}{2},\alpha\uparrow}\bar{\psi}_{-\mathbf{k}+\frac{\mathbf{q}}{2},\beta\downarrow}\psi_{-\mathbf{k}'+\frac{\mathbf{q}}{2},\beta'\downarrow}\psi_{\mathbf{k}'-\frac{\mathbf{q}}{2},\alpha'\uparrow}
    \text{.}
\end{equation}
We write the momentum indices in the ``center of mass" frame of the Cooper pair and decompose the interaction function $V_{\alpha\beta\alpha'\beta'}(\mathbf{k},\mathbf{k}')$
\begin{equation}
    V_{\alpha\beta\alpha'\beta'}(\mathbf{k},\mathbf{k}') = \sum_{i} v_i f^i_{\alpha\beta}(\mathbf{k}) \bar{f}^i_{\alpha'\beta'}(\mathbf{k}')
    \text{,}
\end{equation}
where $i$ indexes some subset of the irreducible representations (irreps) of the crystal symmetry group, and $f^i_{\alpha\beta}(\mathbf{k})$ is the momentum-space form factor associated to irrep $i$. We now perform a Hubbard-Stratonovich transformation to bring this action to a form that is quadratic in $\bar{\psi}$, $\psi$. We decouple in the Cooper channel in terms of order parameters $\bar{\Delta}_{i,\mathbf{q}}$, $\Delta_{i,-\mathbf{q}}$,
\begin{equation}
\begin{split}
    e^{-S_\text{int}[\bar{\psi},\psi]} = \left.\int\mathcal{D}[\bar{\Delta},\Delta] \exp\right\{ &-\int_0^\beta d\tau \sum_{\mathbf{k}\,\alpha\,\beta}\sum_{i\,\mathbf{q}}\left(\bar{\Delta}_{i,\mathbf{q}}\bar{f}^i_{\beta\alpha}(\mathbf{k})\psi_{-\mathbf{k}+\frac{\mathbf{q}}{2},\beta\downarrow}\psi_{\mathbf{k}-\frac{\mathbf{q}}{2},\alpha\uparrow} + \bar{\psi}_{\mathbf{k}-\frac{\mathbf{q}}{2},\alpha\uparrow}\bar{\psi}_{-\mathbf{k}+\frac{\mathbf{q}}{2},\beta\downarrow}\Delta_{i,-\mathbf{q}}f^i_{\alpha\beta}(\mathbf{k})\right) \\ &+ \sum_{i\,\mathbf{q}}\left.\bar{\Delta}_{i,\mathbf{q}}\frac{\beta L^d}{v_i}\Delta_{i,-\mathbf{q}}\right\}\text{.}
\end{split}
\label{HS}
\end{equation}
Here, $\mathbf{q}$ corresponds physically to the \textit{external} momentum of a Cooper pair, in contrast to $\mathbf{k}$, which corresponds to the \textit{internal} momentum. The action can be compactly written using Nambu fields
\begin{equation}
    \Psi_\mathbf{k} \overset{\text{def}}{=}
    \begin{pmatrix}
        \psi_{\mathbf{k},1,\uparrow} \\
        \vdots \\
        \psi_{\mathbf{k},N,\uparrow} \\
        \bar{\psi}_{-\mathbf{k},1,\downarrow} \\
        \vdots \\
        \bar{\psi}_{-\mathbf{k},N,\downarrow} \\        
    \end{pmatrix}
    \text{,}
\end{equation}
in which case the partition function takes the form,
\begin{equation}
    \mathcal{Z} = \int\mathcal{D}[\bar{\Psi},\Psi]\int\mathcal{D}[\bar{\Delta},\Delta] \,\exp\left\{-\int_0^\beta d\tau \int d^d r \left( \bar{\Psi} \hat{\mathcal{G}}^{-1} \Psi - \sum_{i\,\mathbf{q}} \bar{\Delta}_{i,\mathbf{q}}\frac{1}{v_i}\Delta_{i,-\mathbf{q}} \right)\right\}
    \text{,}
\end{equation}
with the inverse Gor'kov Green's function $\hat{\mathcal{G}}^{-1}$. In Matsubara frequency representation, $\Psi(\tau) = \frac{1}{\sqrt{\beta}}\sum_{\omega_n}\Psi_{\omega_n}e^{-i\omega_n\tau}$ and therefore $\partial_\tau \to -i\omega_n$, $\int_0^\beta d\tau \to \beta\sum_{\omega_n}$. (We assume that $\Delta$ has no imaginary time dependence.) This allows writing,
\begin{equation}
    \mathcal{Z} = \int\mathcal{D}[\bar{\Psi},\Psi]\int\mathcal{D}[\bar{\Delta},\Delta] \,\exp\left\{-\sum_{\mathbf{k}\,\mathbf{k}'\,\omega_n} \bar{\Psi}_{\mathbf{k},\omega_n} \left(\hat{\mathcal{G}}^{-1}\right)^{\omega_n}_{\mathbf{k},\mathbf{k}'} \Psi_{\mathbf{k}',\omega_n} + \sum_{i\,\mathbf{q}} \bar{\Delta}_{i,\mathbf{q}}\frac{\beta L^d}{v_i}\Delta_{i,-\mathbf{q}} \right\}
    \text{,}\label{Z matsubara}
\end{equation}
where the matrix elements $\left(\hat{\mathcal{G}}^{-1}\right)^{\omega_n}_{\mathbf{k},\mathbf{k}'}$ of the inverse Green's function operator are given by
\begin{equation}
    \left(\hat{\mathcal{G}}^{-1}\right)^{\omega_n}_{\mathbf{k}+\frac{\mathbf{q}}{2},\mathbf{k}-\frac{\mathbf{q}}{2}} =
    \begin{pmatrix}
        \left(-i\omega_n + \hat{h}_{\mathbf{k}\uparrow}\right)\delta_{\mathbf{q},\mathbf{0}} & \displaystyle \sum_i \Delta_{i,-\mathbf{q}}\, \hat{f}_{\mathbf{k},i} \\
        \displaystyle \sum_i \bar{\Delta}_{i,-\mathbf{q}}\,\hat{f}^\dagger_{\mathbf{k},i} & \left(-i\omega_n - \hat{h}^\top_{-\mathbf{k}\downarrow}\right)\delta_{\mathbf{q},\mathbf{0}}
    \end{pmatrix}
    \text{.}
\end{equation}
Here, we use a condensed notation in which $\hat{h}_{\mathbf{k}\sigma}$ is the matrix with elements $h_{\alpha\beta}^\sigma(\mathbf{k})$, and $\hat{f}_{\mathbf{k},i}$ is the matrix with elements $f_{\alpha\beta}^i(\mathbf{k})$. 
Evaluating the path integral over the Grassmann fields $\bar{\Psi}$, $\Psi$, one obtains
\begin{equation}
    \mathcal{Z} = \int\mathcal{D}[\bar{\Delta},\Delta] \exp\left\{\ln\det \hat{\mathcal{G}}^{-1} + \sum_{i\,\mathbf{q}} \bar{\Delta}_{i,\mathbf{q}}\frac{\beta L^d}{v_i}\Delta_{i,-\mathbf{q}}\right\}
    \text{.}
\end{equation}
with an effective action given by
\begin{equation}
    S_\text{eff}[\bar{\Delta},\Delta] = -\text{tr} \ln \hat{\mathcal{G}}^{-1} - \sum_{i\,\mathbf{q}} \bar{\Delta}_{i,\mathbf{q}}\frac{\beta L^d}{v_i}\Delta_{i,-\mathbf{q}}
    \text{.}
\end{equation}
We expand $S_\text{eff}$ as a functional Taylor series around the saddle point $\Delta=0$,
\begin{equation}
\begin{split}
    S_\text{eff}[\bar{\Delta},\Delta] = \left.S_\text{eff}\right|_{\Delta=0} &+ \sum_{i\,i'\,\mathbf{q}\,\mathbf{q}'} \bar{\Delta}_{i,\mathbf{q}}\left.\frac{\delta^2 S_\text{eff}}{\delta\bar{\Delta}_{i,\mathbf{q}}\delta\Delta_{i',\mathbf{q}'}}\right|_{\Delta=0}\Delta_{i',\mathbf{q}'} \\
    &+ \frac{1}{4}\sum_{ijmn}\bar{\Delta}_{i,\mathbf{0}}\bar{\Delta}_{m,\mathbf{0}}\left.\frac{\delta^4 S_\text{eff}}{\delta\bar{\Delta}_{i,\mathbf{0}}\delta\Delta_{j,\mathbf{0}}\delta\bar{\Delta}_{m,\mathbf{0}}\delta\Delta_{n,\mathbf{0}}}\right|_{\Delta=0}\Delta_{j,\mathbf{0}}\Delta_{n,\mathbf{0}} \\
    &+ \mathcal{O}(|\Delta|^5)
    \text{.}\label{Seff}
\end{split}
\end{equation}
The first-order term vanishes by virtue of $\Delta=0$ being a saddle point. The zeroth-order term is simply the free-electron action, which we can neglect since it just contributes an overall constant. The second-order and fourth-order terms read
\begin{equation}
    S_\text{eff}[\bar{\Delta},\Delta] = \beta L^d \sum_{i\,j\,\mathbf{q}} \bar{\Delta}_{i,\mathbf{q}}\left(-\frac{1}{v_i}\delta_{ij} + \Pi_{ij}^{(2)}(\mathbf{q}) + \frac{1}{4}\sum_{m\,n} \Pi^{(4)}_{ijmn}(\mathbf{0})\,\delta_{\mathbf{q}\mathbf{0}}\bar{\Delta}_{m,\mathbf{0}}\Delta_{n,\mathbf{0}}\right)\Delta_{j,-\mathbf{q}}
    \text{,}\label{Ssp}
\end{equation}
in terms of tensors $\Pi^{(2)}_{ij}(\mathbf{q})$ and $\Pi^{(4)}_{ijmn}(\mathbf{0})$, described in the main text.

\subsection{Second-order terms}

The second-order in $\Delta$ terms (including the gradient terms), are computed in terms of $\Pi^{(2)}_{ii'}(\mathbf{q})$. Comparing Eqs. (\ref{Seff}) and (\ref{Ssp}), we find,
\begin{equation}
    \begin{split}
        \Pi^{(2)}_{ii'}(\mathbf{q}) &= -\frac{1}{\beta L^d}\left.\frac{\delta^2}{\delta\bar{\Delta}_{i,\mathbf{q}}\delta\Delta_{i',\mathbf{q}'}}\text{tr}\ln\hat{\mathcal{G}}^{-1}\right|_{\Delta=0} \\
        &= -\frac{1}{\beta L^d}\left.\frac{\delta}{\delta\bar{\Delta}_{i,\mathbf{q}}}\text{tr}\left(\hat{\mathcal{G}}\frac{\delta\hat{\mathcal{G}}^{-1}}{\delta\Delta_{i',\mathbf{q}'}}\right)\right|_{\Delta=0} \\
        &= \frac{1}{\beta L^d}\,\text{tr}\left(\hat{\mathcal{G}}_0\frac{\delta\hat{\mathcal{G}}_0^{-1}}{\delta\bar{\Delta}_{i,\mathbf{q}}}\hat{\mathcal{G}}_0\frac{\delta\hat{\mathcal{G}}_0^{-1}}{\delta\Delta_{i',\mathbf{q}'}}\right)
        \text{,}
    \end{split}
\end{equation}
where we use the shorthand $\hat{\mathcal{G}}_0 \equiv \left.\hat{\mathcal{G}}\right|_{\Delta=0}$. 
For the honeycomb lattice considered in the main text, the matrix elements of the Green's function $\hat{\mathcal{G}}_0$ are given by
\begin{equation}
\begin{split}
    \left(\hat{\mathcal{G}}_0\right)^{\omega_n}_{\mathbf{k},\mathbf{k}'} &=
    \delta_{\mathbf{k},\mathbf{k}'}
    \begin{pmatrix}
    (-i\omega_n + \hat{h}_{\mathbf{k}\uparrow})^{-1} & 0 \\
    0 & (-i\omega_n - \hat{h}^\top_{-\mathbf{k}\downarrow})^{-1}
    \end{pmatrix}
    \\
    &\equiv \delta_{\mathbf{k},\mathbf{k}'}
    \begin{pmatrix}
    \hat{G}_\mathbf{k}^\uparrow(i\omega_n) & 0 \\
    0 & \hat{G}_\mathbf{k}^\downarrow(i\omega_n)
    \end{pmatrix}
    \text{,}
\end{split}
\end{equation}
where we define $\hat{G}^\uparrow_\mathbf{k}$ and $\hat{G}^\uparrow_\mathbf{k}$ (non-script $G$'s) for later convenience (and we will sometimes suppress the $i\omega_n$ dependence to declutter notation). With these definitions in order, one can carefully compute $\Pi^{(2)}_{ii'}(\mathbf{q})$ making use of the matrix elements of these operators,
\begin{equation}
    \begin{split}
        \Pi^{(2)}_{ii'}(\mathbf{q}) &= \frac{1}{\beta L^d} \sum_{\mathbf{k}_1\,\mathbf{k}_2\,\mathbf{k}_3\,\mathbf{k}_4\,\omega_n} \text{tr}\left\{\left(\hat{\mathcal{G}}_0\right)_{\mathbf{k}_1,\mathbf{k}_2}\left(\frac{\delta\hat{\mathcal{G}}_0^{-1}}{\delta\bar{\Delta}_{i,\mathbf{q}}}\right)_{\mathbf{k}_2,\mathbf{k}_3}\left(\hat{\mathcal{G}}_0\right)_{\mathbf{k}_3,\mathbf{k}_4}\left(\frac{\delta\hat{\mathcal{G}}_0^{-1}}{\delta\Delta_{i',\mathbf{q}'}}\right)_{\mathbf{k}_4,\mathbf{k}_1}\right\} \\
        &= \frac{1}{\beta L^d} \sum_{\{\mathbf{k}_j\}\,\omega_n} \text{tr}\left\{
        \begin{pmatrix}
            \hat{G}^\uparrow_{\mathbf{k}_1} & 0 \\
            0 & \hat{G}^\downarrow_{\mathbf{k}_1}
        \end{pmatrix}
        \delta_{\mathbf{k}_1,\mathbf{k}_2}
        \begin{pmatrix}
        0 & 0 \\
        \hat{f}^\dagger_{\frac{\mathbf{k}_2+\mathbf{k}_3}{2},i} & 0
        \end{pmatrix}
        \delta_{\mathbf{k}_3-\mathbf{k}_2,\mathbf{q}}
        \begin{pmatrix}
            \hat{G}^\uparrow_{\mathbf{k}_3} & 0 \\
            0 & \hat{G}^\downarrow_{\mathbf{k}_3}
        \end{pmatrix}
        \delta_{\mathbf{k}_3,\mathbf{k}_4}
        \begin{pmatrix}
        0 & \hat{f}_{\frac{\mathbf{k}_4+\mathbf{k}_1}{2},i'} \\
        0 & 0
        \end{pmatrix}       
        \delta_{\mathbf{k}_1-\mathbf{k}_4,\mathbf{q}'}\right\} \\
        &= \boxed{\frac{1}{\beta L^d} \sum_{\mathbf{k}\,\omega_n} \text{tr}\left\{\hat{G}^\uparrow_{\mathbf{k}-\frac{\mathbf{q}}{2}}(i\omega_n)\cdot\hat{f}_{\mathbf{k},i'}\cdot\hat{G}^\downarrow_{\mathbf{k}+\frac{\mathbf{q}}{2}}(i\omega_n)\cdot\hat{f}^\dagger_{\mathbf{k},i}\right\}}
        \text{.}\label{trGfGf}
    \end{split}
\end{equation}
Here, $\mathbf{q}'$ must equal $-\mathbf{q}$, as imposed by the Kronecker deltas.

Performing the Matsubara sum in Eq. (\ref{trGfGf}) is made easier by expressing the Green's operators $\hat{G}^\sigma_{\mathbf{k}}(i\omega_n)$ in terms of the Bloch eigenstates $|u^\sigma_{m\mathbf{k}}\rangle$,
\begin{equation}
    \hat{G}^\sigma_{\mathbf{k}}(i\omega_n) = \sum_m \frac{|u^\sigma_{m\mathbf{k}}\rangle\langle u^\sigma_{m\mathbf{k}}|}{-i\omega_n+\epsilon^\sigma_{m\mathbf{k}}}
    \text{.}
\end{equation}
This gives
\begin{align}
    \Pi^{(2)}_{ii'}(\mathbf{q}) &= \frac{1}{\beta L^d}\sum_{mm'}\sum_{\mathbf{k}\,\omega_n}\frac{1}{(-i\omega_n+\epsilon^\downarrow_{m,\mathbf{k}+\frac{\mathbf{q}}{2}})(-i\omega_n+\epsilon^\uparrow_{m',\mathbf{k}-\frac{\mathbf{q}}{2}})}\left\langle u^\downarrow_{m,\mathbf{k}+\frac{\mathbf{q}}{2}} \right| \hat{f}^\dagger_{\mathbf{k},i} \left| u^\uparrow_{m',\mathbf{k}-\frac{\mathbf{q}}{2}} \right\rangle \left\langle u^\uparrow_{m',\mathbf{k}-\frac{\mathbf{q}}{2}} \right| \hat{f}_{\mathbf{k},i'} \left| u^\downarrow_{m,\mathbf{k}+\frac{\mathbf{q}}{2}} \right\rangle \notag\\
    &= \frac{1}{\beta L^d}\sum_{mm'}\sum_{\mathbf{k}} \frac{n_\text{F}(\epsilon^\downarrow_{m,\mathbf{k}+\frac{\mathbf{q}}{2}}) - n_\text{F}(\epsilon^\uparrow_{m',\mathbf{k}-\frac{\mathbf{q}}{2}})}{\epsilon^\downarrow_{m,\mathbf{k}+\frac{\mathbf{q}}{2}}-\epsilon^\downarrow_{m',\mathbf{k}-\frac{\mathbf{q}}{2}}} \left\langle u^\downarrow_{m,\mathbf{k}+\frac{\mathbf{q}}{2}} \right| \hat{f}^\dagger_{\mathbf{k},i} \left| u^\uparrow_{m',\mathbf{k}-\frac{\mathbf{q}}{2}} \right\rangle \left\langle u^\uparrow_{m',\mathbf{k}-\frac{\mathbf{q}}{2}} \right| \hat{f}_{\mathbf{k},i'} \left| u^\downarrow_{m,\mathbf{k}+\frac{\mathbf{q}}{2}} \right\rangle
    ,
\end{align}
where $n_\text{F}(x) = (1 + \exp(\beta x))^{-1}$ is the Fermi-Dirac distribution function. If the system respects time-reversal symmetry, then $\hat{h}^\uparrow_\mathbf{k} = \left(\hat{h}^\downarrow_{-\mathbf{k}}\right)^*$, and $\Pi^{(2)}_{ii'}(\mathbf{q})$ becomes,
\begin{equation}
    \Pi^{(2)}_{ii'}(\mathbf{q}) = \frac{1}{\beta L^d}\sum_{mm'}\sum_{\mathbf{k}} \frac{n_\text{F}(\epsilon_{m,\mathbf{k}+\frac{\mathbf{q}}{2}}) - n_\text{F}(-\epsilon_{m',\mathbf{k}-\frac{\mathbf{q}}{2}})}{\epsilon_{m,\mathbf{k}+\frac{\mathbf{q}}{2}}+\epsilon_{m',\mathbf{k}-\frac{\mathbf{q}}{2}}} \left\langle u_{m,\mathbf{k}+\frac{\mathbf{q}}{2}} \right| \hat{f}^\dagger_{\mathbf{k},i} \left| u_{m',\mathbf{k}-\frac{\mathbf{q}}{2}} \right\rangle \left\langle u_{m',\mathbf{k}-\frac{\mathbf{q}}{2}} \right| \hat{f}_{\mathbf{k},i'} \left| u_{m,\mathbf{k}+\frac{\mathbf{q}}{2}} \right\rangle
    .
\end{equation}

\subsection{Fourth-order terms}

Following steps analogous to the previous section, we find that $\Pi_{ijmn}^{(4)}(\mathbf{0})$ equals the following (assuming time-reversal symmetry of the Bloch Hamiltonian),
\begin{equation}
\begin{split}
    \Pi^{(4)}_{ijmn}(\mathbf{0}) &\equiv \left.\frac{\delta^4 S_\text{eff}}{\delta\bar{\Delta}_{i,\mathbf{0}}\delta\Delta_{j,\mathbf{0}}\delta\bar{\Delta}_{m,\mathbf{0}}\delta\Delta_{n,\mathbf{0}}}\right|_{\Delta=0} \\
    &=\frac{1}{\beta L^d} \sum_{\mathbf{k}\,\omega_n} \text{tr}\left\{\hat{G}^\uparrow_{\mathbf{k}}(i\omega_n)\cdot\hat{f}_{\mathbf{k},n}\cdot\hat{G}^\downarrow_{\mathbf{k}}(i\omega_n)\cdot\hat{f}^\dagger_{\mathbf{k},m}\cdot\hat{G}^\uparrow_{\mathbf{k}}(i\omega_n)\cdot\hat{f}_{\mathbf{k},j}\cdot\hat{G}^\downarrow_{\mathbf{k}}(i\omega_n)\cdot\hat{f}^\dagger_{\mathbf{k},i}\right\}.
\end{split}
\end{equation}
Writing this in terms of Bloch eigenstates (assuming time-reversal symmetry, for simplcity), we have,
\begin{equation}
    \Pi^{(4)}_{i_1\hdots i_4} = \frac{1}{\beta L^d}\sum_{m_1\hdots m_4}\sum_{\mathbf{k}\,\omega_n} \frac{\langle u_{m_1\mathbf{k}}|\hat{f}_{\mathbf{k},i_1}|u_{m_2\mathbf{k}}\rangle\langle u_{m_2\mathbf{k}} | \hat{f}^\dagger_{\mathbf{k},i_2}|u_{m_3\mathbf{k}}\rangle\langle u_{m_3\mathbf{k}}|\hat{f}_{i_3\mathbf{k}}|u_{m_4\mathbf{k}}\rangle\langle u_{m_4\mathbf{k}}|\hat{f}^\dag_{i_4\mathbf{k}}|u_{m_1\mathbf{k}}\rangle}{(-i\omega_n + \epsilon_{m_1\mathbf{k}})(+i\omega_n-\epsilon_{m_2\mathbf{k}})(-i\omega_n+\epsilon_{m_3\mathbf{k}})(+i\omega_n+\epsilon_{m_4\mathbf{k}})}.
\end{equation}
The Matsubara sum here requires a bit more care to compute,
\begin{equation}
\begin{split}
    &\sum_{i\omega_n}\frac{1}{(-i\omega_n + \epsilon_{m_1\mathbf{k}})(+i\omega_n-\epsilon_{m_2\mathbf{k}})(-i\omega_n+\epsilon_{m_3\mathbf{k}})(+i\omega_n+\epsilon_{m_4\mathbf{k}})} \\
    &= \frac{1}{\epsilon_{m_1\mathbf{k}}-\epsilon_{m_3\mathbf{k}}}\left[\frac{n_\text{F}(\epsilon_{m_1\mathbf{k}})}{(\epsilon_{m_1\mathbf{k}}+\epsilon_{m_2\mathbf{k}})(\epsilon_{m_1\mathbf{k}}+\epsilon_{m_4\mathbf{k}})}-\frac{n_\text{F}(\epsilon_{m_3\mathbf{k}})}{(\epsilon_{m_3\mathbf{k}}+\epsilon_{m_2\mathbf{k}})(\epsilon_{m_3\mathbf{k}}+\epsilon_{m_4\mathbf{k}})}\right] \\
    &- \frac{1}{\epsilon_{m_2\mathbf{k}}-\epsilon_{m_4\mathbf{k}}}\left[\frac{n_\text{F}(-\epsilon_{m_2\mathbf{k}})}{(\epsilon_{m_2\mathbf{k}}+\epsilon_{m_1\mathbf{k}})(\epsilon_{m_2\mathbf{k}}+\epsilon_{m_3\mathbf{k}})}-\frac{n_\text{F}(-\epsilon_{m_4\mathbf{k}})}{(\epsilon_{m_4\mathbf{k}}+\epsilon_{m_1\mathbf{k}})(\epsilon_{m_4\mathbf{k}}+\epsilon_{m_3\mathbf{k}})}\right] \\
    &\equiv Q(\epsilon_{m_1\mathbf{k}},\epsilon_{m_2\mathbf{k}},\epsilon_{m_3\mathbf{k}},\epsilon_{m_4\mathbf{k}}) + Q(-\epsilon_{m_2\mathbf{k}},-\epsilon_{m_1\mathbf{k}},-\epsilon_{m_4\mathbf{k}},-\epsilon_{m_3\mathbf{k}}),
\end{split}
\end{equation}
where for convenience we define a function $Q(\epsilon_1,\epsilon_2,\epsilon_3,\epsilon_4)$,
\begin{equation}
    Q(\epsilon_1,\epsilon_2,\epsilon_3,\epsilon_4) \overset{\text{def}}{=} \frac{1}{\epsilon_{1}-\epsilon_{3}}\left[\frac{n_\text{F}(\epsilon_{1})}{(\epsilon_{1}+\epsilon_{2})(\epsilon_{1}+\epsilon_{4})}-\frac{n_\text{F}(\epsilon_{3})}{(\epsilon_{3}+\epsilon_{2})(\epsilon_{3}+\epsilon_{4})}\right]
    \text{.}
\end{equation}
$Q$ has a well-defined limit as $\epsilon_1$ approaches $\epsilon_3$
\begin{equation}
    Q_0(\epsilon_2,\epsilon_3,\epsilon_4) \overset{\text{def}}{=} \lim_{\epsilon_1\to\epsilon_3} Q(\epsilon_1,\epsilon_2,\epsilon_3,\epsilon_4) = \frac{n'_\text{F}(\epsilon_3)}{(\epsilon_3+\epsilon_2)(\epsilon_3+\epsilon_4)} - n_\text{F}(\epsilon_3)\left[\frac{1}{(\epsilon_3+\epsilon_2)(\epsilon_3+\epsilon_4)^2}+\frac{1}{(\epsilon_3+\epsilon_2)^2(\epsilon_3+\epsilon_4)}\right],
\end{equation}
where $n_\text{F}'(x)$ is the derivative of the Fermi-Dirac distribution function. The combined formal expression for $\Pi^{(4)}_{i_1\hdots i_4}$ reads
\begin{equation}
\begin{split}
    \Pi^{(4)}_{i_1\hdots i_4} &= \frac{1}{\beta L^d}\sum_{m_1\hdots m_4}\sum_{\mathbf{k}} \left[Q(\epsilon_{m_1\mathbf{k}},\epsilon_{m_2\mathbf{k}},\epsilon_{m_3\mathbf{k}},\epsilon_{m_4\mathbf{k}}) + Q(-\epsilon_{m_2\mathbf{k}},-\epsilon_{m_1\mathbf{k}},-\epsilon_{m_4\mathbf{k}},-\epsilon_{m_3\mathbf{k}})\right] \\
    &\,\,\,\,\,\,\,\,\,\,\,\,\,\,\,\,\,\,\,\,\,\,\,\,\,\,\,\,\,\,\,\,\,\,\,\,\,\,\,\,\,\,\,\times\langle u_{m_1\mathbf{k}}|\hat{f}_{\mathbf{k},i_1}|u_{m_2\mathbf{k}}\rangle\langle u_{m_2\mathbf{k}} | \hat{f}^\dagger_{\mathbf{k},i_2}|u_{m_3\mathbf{k}}\rangle\langle u_{m_3\mathbf{k}}|\hat{f}_{i_3\mathbf{k}}|u_{m_4\mathbf{k}}\rangle\langle u_{m_4\mathbf{k}}|\hat{f}^\dag_{i_4\mathbf{k}}|u_{m_1\mathbf{k}}\rangle
    ,
\end{split}
\end{equation}
where it is understood that the proper limit $Q_0$ must be explicitly implemented whenever the first and third argument of $Q$ are equal.